\documentclass[12pt]{article}
\usepackage{epstopdf}
\usepackage{fullpage}
\usepackage[latin1]{inputenc}
\usepackage[T1]{fontenc}
\usepackage{ae,aecompl}
\usepackage{amsmath,amsfonts}

\usepackage{booktabs}

\newenvironment{pf}{\noindent\textbf{Proof.}\quad}{\hfill{$\Box$}}
\usepackage[pdftex]{hyperref}
\usepackage[dvips]{graphicx}
\usepackage[usenames,dvipsnames]{color}
\usepackage{epsfig}
\usepackage{rotating}
\usepackage{amssymb}
\usepackage[latin1]{inputenc}
\usepackage{verbatim}
\newtheorem{lem}{Lemma}
\newtheorem{df}{Definition}
\newtheorem{thm}{Theorem}

\newtheorem{conj}{Conjecture}

\newtheorem{cor}{Corollary}
\newcommand{\F}{\mathbb{F}}
\newcommand{\Z}{\mathbb{Z}}

\newcommand{\beg}{\begin{equation}}
\newcommand{\eeg}{\end{equation}}
\newcommand{\bearray}{\begin{eqnarray}}
\newcommand{\eearray}{\end{eqnarray}}

\newcommand{\m}{\mbox}
\newcommand{\mf}{\hspace{5mm} \mbox{ }}
\newcommand{\mz}{\hspace{2mm} \mbox{ }}
\newcommand{\bite}{\begin{itemize}}
\newcommand{\eite}{\end{itemize}}
\newcommand{\al}{\alpha}

\begin{document}

\title{A complementary construction using mutually unbiased bases}
\author{Gaofei Wu and Matthew G. Parker
\thanks{G. Wu is with the State Key Laboratory of Integrated Service Networks,
Xidian University, Xi'an, 710071, China. He is a visiting PhD student
(Sep. 2012 -- Aug. 2014) in the Department of Informatics,
University of Bergen, Norway: gaofei.wu@student.uib.no, M. G. Parker is with
the Department of Informatics,
University of Bergen, Norway: matthew@ii.uib.no.}
}

\date{\today}

\maketitle

\begin{abstract}
We propose a construction for
complementary sets of arrays that exploits a set of mutually-unbiased bases (a MUB). In particular we
present, in detail, the construction for complementary pairs that is seeded by a MUB of
dimension 2, where we
enumerate the arrays and the corresponding set of complementary sequences obtained from the
arrays by projection. We also
sketch an algorithm to uniquely generate these sequences.
The pairwise squared inner-product
of members of the sequence set is shown to be $\frac{1}{2}$.
Moreover, a subset of the set can be viewed as a codebook that asymptotically achieves
$\sqrt{\frac{3}{2}}$ times the Welch bound.
\end{abstract}

\section{Introduction}
Sequences with relatively flat Fourier spectra are of central importance for many communications
systems such as spread-spectrum, and are also used to probe structures in
the context of measurement and detection.
It is often the case that a set of such sequences is required, where family members are pairwise
distinguishable - in the context of communication each user may be assigned a different
sequence from the set, and in the context of measurement the pairwise distinguishability implies
that, when probing a structure, each sequence in the set contributes useful information to the
overall measurement or detection problem. One recent communication application is to
orthogonal frequency-division multiplexing (OFDM), which is a communication
 technique used in several wireless communication standards such
  as IEEE 802.16 Mobile WiMAX.  A major problem with
    OFDM is the large peak-to-average power ratio (PAPR) of uncoded OFDM time signals (i.e.
    the signals do not have relatively flat inverse Fourier spectra).
{\em Complementary sequences} \cite{golay,Shap,Budisin} are sets of sequences
that have out-of-phase aperiodic autocorrelations
that sum to zero. This implies that they have very flat Fourier
spectra - the Fourier transform of each sequence satisfies a PAPR upper bound of $2.0$,
which is very low, and Davis and Jedwab \cite{davis} showed, in the context of OFDM,
 how to construct `standard'
     $2^h$-ary complementary sequences of length $2^n$, comprising
     second-order cosets of generalized
   first-order Reed-Muller codes $RM_{2^h}(1,n)$. As they are members of $RM_{2^h}(1,n)$, the
   sequences have good pairwise distinguishability.
The work was subsequently extended by \cite{paterson,schmidt,schmidt2} and by numerous other authors \cite{Zilong}.
\cite{parker,parker3,Mats,fiedler1,JP,parker2011} show that the complementary set
    construction is primarily an array construction, where sequence sets are obtained by
    considering suitable projections of the arrays.
It is desirable to propose complementary constructions that significantly improve set size without
greatly compromising the upper bound on PAPR or the pairwise distinguishability.

In this paper we show that the problem of
construction of large sets of complementary sequences
with good pairwise distinguishability is naturally solved by seeding the recursive
construction with optimal {\em mutually-unbiased bases} (MUBs) \cite{Schwing,ivan81,Ruskai}.
By way of example, we construct a set of complementary array pairs over
the alphabet ${\cal A} = \{0,1,i,-1,-i\}$ (up to normalisation of the sequence),
whose sequence projections
can be viewed as a superset of the standard quaternary complementary sequences
of \cite{davis}, where our construction enlarges set size without compromising
PAPR or pairwise distinguishability,
at the cost of adding `$0$' to the QPSK alphabet - this construction exploits
an optimal MUB of dimension 2. The exact number of arrays that we construct is determined.
An algorithm for generating
all unique sequence projections from these arrays is then sketched out and an implementation of this
algorithm allows us to compute
the number of these sequences. These computational results then help us to theoretically
establish corresponding enumeration formulae, where we are guided in our theoretical development
by related enumerations that we found in the Online Encyclopaedia of Integer Sequences \cite{Sloane}.
The construction generates a set of complementary
sequences which is a relatively large superset of the complementary sequences obtained
in \cite{davis}, but we show that the magnitude of the
pairwise inner product between members of the set remains at
$\frac{1}{\sqrt{2}}$, the same as for the set in \cite{davis}.

This paper can also be seen as a generalisation of the construction in
\cite{bparker09}, where the
authors proposed a family of complementary sequences over  $\{0,1,-1\}$.
Moreover the paper can be seen as an explicit consequence of the more general
principles for the construction of complementary sets, as described in \cite{parker2011}.
There exist related matrix-based approaches to complementary sequence design in the literature.
For instance, the complete complementary code approach of \cite{Suehiro} and the
paraunitary matrix approach to filter banks for complementary sequences, as discussed in
\cite{BudSpas,BudQAM} - the word `paraunitary' refers to a matrix polynomial in $Z^{-1}$ that is unitary
when $|Z| = 1$, and the approach of these papers is clearly similar to our own.
In particular, the recent work in \cite{BudQAM} makes extensive
recursive use of paraunitary matrices to generate QAM complementary sequence pairs.

A subset of our sequence set can be viewed as a codebook, where
the magnitude of the pairwise inner product between codewords in the codebook approaches
$\sqrt{\frac{3}{2}}$ times the Welch bound as length increases \cite{welch,sarwate}.

After some preliminaries in section \ref{pre},
we introduce our main construction in section \ref{maincon}. To begin with we develop the complementary
construction in a general way, for complementary sets, so as to emphasise that we can seed with any
MUB of any dimension. But, for the general case,
it remains open to develop formulae for the size of the array and sequence sets, and for the
magnitude of the pairwise inner product between members of the sets. So, for this paper, we only
develop in detail the case where we seed our
construction with an optimal MUB of dimension 2, i.e. we
construct complementary pairs. Nevertheless this is an important case, and it serves to illustrate
more general principles.
In section \ref{enum} we give the
exact enumeration of the complementary arrays and sequences we construct, as well as sketch
an algorithm to generate the sequences uniquely.  The maximum pairwise  inner product between
sequences in our set is determined in section \ref{distance}. In section
 \ref{codebooks}, we give our codebook construction. Section \ref{conclusion} concludes with some open
 problems.

\section{Preliminaries}\label{pre}

\subsection{Mutually unbiased bases}
\label{MUB}

Denote the magnitude of the normalised
pairwise inner product of two equal-length complex vectors, $u$ and $v$, by
$$ \Delta(u,v) = \frac{|\langle u,v \rangle|}{|u|\cdot |v|}. $$

\vspace{2mm}

A pair of bases $u_0,\cdots, u_{\delta-1}$ and $v_0,\cdots, v_{\delta-1}$ in ${\mathbb C}^{\delta}$
is said to be
{\em mutually unbiased} if they are both orthonormal and there is a constant $a$ such that
$\Delta^2(u_i,v_j) = |\langle u_i,v_j\rangle|^2=a$, $\forall i,j$. A set of bases is then called a
set of mutually unbiased bases (MUB) if any pair of them is mutually unbiased.
It is known that a MUB contains at most $\delta+1$ bases in
${\mathbb C}^{\delta}$, in which case the MUB is referred to as an {\em optimal} MUB.
Such optimal MUBs exist
if $\delta$ is a prime power \cite{ivan81}, in which case $a = \frac{1}{\delta}$.
We refer to an optimal MUB as ${\cal M}_{\delta}$
\begin{footnote}{Up to trivial symmetries and, for a fixed $\delta$,
there may be more than one choice of ${\cal M}_{\delta}$.}
\end{footnote}.
Specifically, in this paper, we focus on a particular matrix
form of ${\cal M}_2$, where the 3 bases are the rows of the 3 unitary
matrices, $I$, $H$, and $N$, where
$I = \left ( \begin{tiny} \begin{array}{cc} 1 & 0 \\ 0 & 1 \end{array} \end{tiny} \right )$,
$H = \frac{1}{\sqrt{2}}\left ( \begin{tiny} \begin{array}{cc} 1 & 1 \\ 1 & -1 \end{array} \end{tiny} \right )$,
and
$N = \frac{1}{\sqrt{2}}\left ( \begin{tiny} \begin{array}{cc} 1 & i \\ 1 & -i \end{array} \end{tiny} \right )$,
where $i = \sqrt{-1}$.

\vspace{2mm}

The properties of a matrix MUB remain unchanged with respect to
row/column re-ordering and multiplication of a row/column by a unity phase shift, as
summarised by the following equivalence relationship between two
$S \times S$ unitary complex matrices,
$M$ and $M'$:
$$ M' = O_{\theta}P_{\theta}MP_{\gamma}O_{\gamma}, $$
where
$O_{\gamma}$ and $O_{\theta}$ are diagonal unitary matrices, and
$P_{\gamma}$ and $P_{\theta}$ are permutation matrices.
However there is no distance between two vectors whose elements differ only by a constant
phase shift, so to ensure that $\Delta$ for our constructed sequence set is nonzero, we
force $O_{\theta} = O_{\gamma} = I$ and, therefore, only use the equivalence
\beg M' = P_{\theta}MP_{\gamma}. \label{equiv} \eeg

\vspace{2mm}

In this paper we also make use of the Pauli matrices
$X = \left ( \begin{tiny} \begin{array}{cc} 0 & 1 \\ 1 & 0 \end{array} \end{tiny} \right )$
and $Z = \left ( \begin{tiny} \begin{array}{cc} 1 & 0 \\ 0 & -1 \end{array} \end{tiny} \right )$.

\subsection{Complementary arrays and sequences}
Let $F(z) = (F_0(z),F_1(z),\ldots,F_{S-1}(z))^T$ be a length $S$ vector of
polynomials, where $F_k(z) = \sum_{i=0}^{d_k-1} F_{k,i}z^i$, $F_{k,i} \in {\mathbb C}$,
$\forall i$, is a complex polynomial of degree $d_k-1$, $0 \le k < S$.
We also associate with and refer to $F_k(z)$ as the length $d_k$ sequence
$(F_{k,0},F_{k,1},\ldots,F_{k,d_k-1})$.
The {\em aperiodic autocorrelation} of $F_k$ is
given by the coefficients of $F_k(z)F_k^*(z^{-1})$, where $*$ means complex conjugate.
Let
$$ \lambda_F(z) = \langle F(z),F(z) \rangle = F^{\dag}(z^{-1})F(z), $$
be the inner-product of $F$ with
itself, where $\dag$ means `transpose-conjugate'
\begin{footnote}{
Comparing $F_k(z)F_k^*(z^{-1})$ with $F^{\dag}(z^{-1})F(z)$, we see that, whilst $z^{-1}$ is on the
right for the former it is on the left for the latter. This is simply because $F$ is a $S \times 1$ vector
 - there is no deeper meaning.
}\end{footnote}
.
For $S \ge 2$, we desire to find $S$ degree $d-1$ polynomials, $F_k(z)$,
such that $\lambda_F(z) = \lambda_F$, a constant independent of $z$, in which case
the set ${\cal F}(z) = \{F_0(z),F_1(z),\ldots,F_{S-1}(z)\}$ is called
a size $S$ {\em complementary set} of length $d$ sequences.

\vspace{2mm}

\noindent {\bf Example: } Let $S = 2$, $F_0(z) = 1 + z + z^2 - z^3$, $F_1(z) = 1 + z - z^2 + z^3$.
Then $\langle F(z),F(z) \rangle = \lambda_F(z)
 = (-z^{-3} + z^{-1} + 4 + z - z^3) + (z^{-3} - z^{-1} + 4 - z + z^3) = 8$, so
${\cal F}(z) = \{F_0(z),F_1(z)\}$ is a size $2$ complementary set of length 4 sequences,
where the sequences are
$(1,1,1,-1)$ and $(1,1,-1,1)$.

\vspace{2mm}

The complementary set property can be interpreted in the Fourier domain by evaluating $z$ on
the unit circle. If ${\cal F}$ is a complementary set of size $S$, then
$$ \lambda_F = F^{\dag}(\al^{-1})F(\al), \mf |\al| = 1. $$
It follows that
\beg F_k(\al)F_k^*(\al^{-1}) \le \lambda_F, \mf |\al| = 1, 0 \le k < S. \label{SpectralBound} \eeg
(\ref{SpectralBound}) states that the Fourier power spectrum of each sequence $F_k$, $0 \le k < S$,
is upper-bounded by $\lambda_F$. If $\|F_k\|^2 = \sum_{0 \le i < n} F_{k,i}F_{k,i}^* = 1$,
$0 \le k < S$, i.e. if each of the $S$ sequences
has its power normalised to 1, then $\lambda_F = S$ and we say that the
{\em peak-to-average power ratio} (PAPR)
of each sequence in ${\cal F}$ is upper-bounded by $S$.

We can generalize further by replacing $z$ with ${\bf z} = (z_0,z_1,\ldots,z_{n-1})$, where
$F_k({\bf z})$ is of degree $d_{k,j}-1$ in variable $z_j$, $0 \le j < n$, in which case $F_k({\bf z})$
is associated with and referred to as an $n$-dimensional
$d_{k,0} \times d_{k,1} \times \ldots d_{k,n-1}$ complex array. The aperiodic autocorrelation of $F_k$ is
given by the coefficients of $F_k({\bf z})F_k^*({\bf z}^{-1})$, where
${\bf z}^{-1} = (z_0^{-1},z_1^{-1},\ldots,z_{n-1}^{-1})$ and, as before, ${\cal F}({\bf z})$ is a
complementary set of size $S$ if
$$ \lambda_F = \lambda_F({\bf z}) = \langle F({\bf z}),F({\bf z}) \rangle
 = F^{\dag}({\bf z}^{-1})F({\bf z}). $$

\vspace{2mm}

\noindent {\bf Example: } Let $S = 2$, $F_0({\bf z}) = 1 + z_0 + z_1 - z_0z_1$, $F_1({\bf z}) = 1 + z_0 - z_1 + z_0z_1$.
Then $\langle F({\bf z}),F({\bf z}) \rangle = \lambda_F({\bf z})
 = (-z_0^{-1}z_1^{-1} + z_0z_1^{-1} + 4 + z_0^{-1}z_1 - z_0z_1)
 + (z_0^{-1}z_1^{-1} - z_0z_1^{-1} + 4 - z_0^{-1}z_1 + z_0z_1) = 8$, so
${\cal F}({\bf z}) = \{F_0({\bf z}),F_1({\bf z})\}$ is a size $2$ complementary set
of $2 \times 2$ arrays, where the arrays are
$\left ( \begin{tiny} \begin{array}{rr} 1 & 1 \\ 1 & -1 \end{array} \end{tiny} \right)$ and
$\left ( \begin{tiny} \begin{array}{rr} 1 & 1 \\ -1 & 1 \end{array} \end{tiny} \right)$.

\vspace{2mm}

Section \ref{maincon} introduces the construction for complementary sets, i.e. general $S$, and this
construction can be seeded with an optimal MUB, ${\cal M}_{\delta}$,
where $\delta = S$. But we subsequently
only develop formulae for the case where $\delta = S = 2$ so, for that case,
$d_{k,j} = d_j = 2$, $0 \le j < n$, $\forall k$,
and therefore
$F_k({\bf z}) \in ({\mathbb C}^2)^{\otimes n}$, i.e. the coefficients of $F_k({\bf z})$ form an array $\in ({\mathbb C}^2)^{\otimes n}$.
We shall then recursively
construct a set of complementary pairs of arrays, ${\cal F}({\bf z})$. The set of distinct complementary
arrays obtained from ${\cal F}({\bf z})$ is called ${\cal B}_n$.
Then, by applying projections
$z_i = z^{2^{\pi(i)}}$, $0 \le i < n$,
over all permutations, $\pi$, in the symmetric group
${\cal S}_n$, we shall construct, from ${\cal F}({\bf z})$,
a set of complementary pairs of sequences, ${\cal F}(z)$.
The set of complementary sequences obtained from ${\cal F}(z)$ is called ${\cal B}_{\downarrow,n}$.
We shall compute values for
$|{\cal B}_n|$ and $|{\cal B}_{\downarrow,n}|$ in terms of $n$, and then develop theoretical
formulae for these two parameters that agree with our computations.
We shall also determine, theoretically, that,
when seeded with ${\cal M}_2 = \{I,H,N\}$,
$$ \Delta^2({\cal B}_{\downarrow,n}) = \m{ max}\{\Delta^2(u,v) | u \ne v, u,v \in {\cal B}_{\downarrow,n}\} = \frac{1}{2}. $$

\vspace{2mm}

\noindent {\bf Example: } The array $F_k({\bf z}) = 1 + z_0 - z_1 + z_0z_1$ can be projected
down to the sequence $F_k(z) = 1 + z - z^2 + z^3$ by the assignment $z_1 = z^2$, $z_0 = z$,
and to $F_k(z) = 1 - z + z^2 + z^3$ by the assignment $z_0 = z^2$, $z_1 = z$.

\vspace{3mm}

As well as expressing our arrays and sequences as the coefficients of
polynomials $F_k({\bf z})$ and $F_k(z)$, respectively, we can, for the case $S = 2$, and where
we seed our construction with ${\cal M}_2 = \{I,H,N\}$,
further express them as
generalized Boolean functions, $f_k({\bf x}) : \F_2^n \rightarrow {\cal A}$, where
${\cal A} = \{0,1,i,-1,-i\}$, $i = \sqrt{-1}$, and
$F_k({\bf z}) = c\sum_{{\bf x} \in \F_2^n} f_k({\bf x}){\bf z}^{\bf x}$, where $c$ is some
normalising constant, chosen so that the associated array, $F_k$, satisfies
$\|F_k({\bf z})\|^2 = \sum_{0 \le i < n} F_{k,i}F_{k,i}^* = 1$. When we refer to $f_k({\bf x})$
as an array or sequence, we mean that the $2^n$ elements of the array or sequence are the $2^n$ evaluations of $f_k({\bf x})$
for ${\bf x} \in \F_2^n$. These elements are also
the $2^n$ coefficients of $F_k({\bf z})$ or $F_k(z)$, respectively.
The interpretation of $f_k({\bf x})$ as one of a set of $n!$
sequence projections, depending on
$\pi \in {\cal S}_n$, as opposed to the parent array, is implicitly made and
should be clear from the context of the discussion.

\vspace{2mm}

\noindent {\bf Example: } The array
$F_k({\bf z}) = c(1 + z_0 + iz_1z_2 - iz_0z_1z_2)$, where $c = \frac{1}{2}$,
may be represented by
$f_k({\bf x}) = (x_1 + x_2 + 1)i^{2x_0x_2 + x_2}$, and $f_k$ may also refer, via
projection, to one
of $3! = 6$ sequences, e.g. with $z_2 = z$, $z_0 = z^2$, $z_1 = z^4$, we project
to $F_k(z) = c(1 + z^2 + iz^5 - iz^7)$, which is then one of the 6 sequences represented by $f_k$
\begin{footnote}{More accurately, we should write $f_k(\pi({\bf x}))$ to indicate one of 6 possible
permutations but, to reduce notation, we make such a mapping implicit in this paper.}
\end{footnote}.

\vspace{2mm}

In section 3, after presenting the general complementary set construction,
we explicitly seed the construction with ${\cal M}_2 = \{I,H,N\}$ and, thereby,
recursively construct a set of complementary arrays, ${\cal B}_n$,
and project the arrays
down to a set of sequences, ${\cal B}_{\downarrow,n}$, where the sequence
PAPR upper bound of $S = \delta = 2$ follows from the unitarity of $I$, $H$, and $N$, and where the value
of $\Delta^2({\cal B}_{\downarrow,n}) = \frac{1}{2}$ follows from the value of $\Delta$ for ${\cal M}_2$.

\section{The complementary set construction}\label{maincon}
\subsection{The general construction}
Let ${\bf w} = (w_0,w_1,\ldots,w_{m-1}) \in {\mathbb C}^m$, and
${\bf y} = (y_0,y_1,\ldots,y_{m'-1}) \in {\mathbb C}^{m'}$,
be disjoint vectors of $m$ and $m'$ complex variables, respectively, and let
${\bf z} = {\bf w} \cup {\bf y} \in {\mathbb C}^{m + m'}$ be the vector of variables
formed from the union of variables in ${\bf w}$ and ${\bf y}$, with some ordering on the variables.
Let $F_j({\bf w}) : {\mathbb C}^m \rightarrow {\mathbb C}$ be of degree $d_j - 1$ in variable $w_j$, $0 \le j < m$.
Let
$F({\bf w}) = (F_0({\bf w}),F_1({\bf w}),\ldots,F_{S-1}({\bf w}))^T$, and let ${\cal{U}}({\bf y}) = (u_{ij}({\bf y}), 0 \le i,j < S)$
be an $S \times S$ unnormalised unitary matrix with
elements being complex polynomials $u_{ij}({\bf y}) : {\mathbb C}^{m'} \rightarrow {\mathbb C}$.
By unnormalised unitary we mean that
${\cal{U}}({\bf y}){\cal{U}}^{\dag}({\bf y}) = \lambda_{\cal U}({\bf y})I$, where
$I$ is the $S \times S$ identity matrix.

Let $\lambda_F({\bf w}) = \sum_{k=0}^{S-1} F_k({\bf w}){F_k}^*({\bf w^{-1}})$, and define $F'({\bf z})$ by
\beg F'({\bf z}) = {\cal{U}}({\bf y})F({\bf w}). \label{GenComp} \eeg

It follows from the unitarity of ${\cal{U}}$, and from (\ref{GenComp}), that
\beg \lambda_{F'}({\bf z}) = \sum_{k=0}^{S-1} F'_k({\bf z}){F'_k}^*({\bf z^{-1}})
 = \lambda_{\cal U}({\bf y})\lambda_F({\bf w}). \label{lambda} \eeg

If $\lambda_F({\bf w}) = c_F$, a constant, independent of ${\bf w}$, and
$\lambda_{\cal U}({\bf y}) = c_{\cal U}$, a constant, independent of ${\bf y}$ then
$\lambda_{F'}({\bf z}) = c_{F'} = c_{\cal U}c_F$ is a constant, independent of ${\bf z}$. If so, then
(\ref{GenComp})
defines a step in the construction of generalised $|{\bf z}|$-dimensional complementary array sets
of size $S$ from $|{\bf w}|$-dimensional complementary array sets of size $S$, and (\ref{lambda})
characterises the complementary property that the sum of the $S$ aperiodic array autocorrelations,
$F'_k({\bf z}){F'_k}^*({\bf z^{-1}})$, is a constant, independent of ${\bf z}$.
In this paper we only consider the case where
$w_i \ne y_j$, $\forall i,j$, but complementarity holds even when this is not true.

\vspace{2mm}

\noindent {\bf Example: } Let $F(w) = (1+w, 1-w)^T$ and
${\cal U}(y) = \left ( \begin{tiny} \begin{array}{rr} 1 & y \\ 1 & -y \end{array} \end{tiny} \right )$.
Then $F'({\bf z}) = F'(y,w) = {\cal U}(y)F(w) = (1 + w + y - wy, 1 + w - y + wy)^T$. As
$\lambda_F(w) = F(w^{-1})^{\dag}F(w) = 4$, and $\lambda_{\cal U}(y) = 2$, then
$\lambda_{F'}(z) = 8$. So the arrays comprising the coefficients of $1+w+y-wy$ and $1+w-y+wy$ are a
complementary set of size $S = 2$, i.e. a complementary pair.

\vspace{2mm}

We can recurse (\ref{GenComp}). With notational changes,
the $j$'th recursive step of (\ref{GenComp}) is described by,
\beg F_j({\bf z}_j) = {\cal{U}}_j({\bf y}_j)F_{j-1}({\bf z}_{j-1}), \label{GenComp1} \eeg
where ${\bf y}_j = (z_{\mu_j},z_{\mu_j+1},\ldots,z_{\mu_j + m_j-1})$,
${\bf z}_j = (z_0,z_1,\ldots,z_{\mu_j + m_j-1})$, $\mu_j = \sum_{i=0}^{j-1} m_j$, $\mu_0 = 0$,
$F_j({\bf z}_j) = (F_{j,0}({\bf z}_j),F_{j,1}({\bf z}_j),\ldots,F_{j,S-1}({\bf z}_j))^T$,
and $F_{-1} = \frac{1}{\sqrt{S}}(1,1,\ldots,1)$. (\ref{GenComp1}) is a very general recursive equation for the construction
of complementary sets of arrays of size $S$.

\vspace{3mm}

We argue, in this paper, that it is natural, for (\ref{GenComp1}), to
choose ${\cal{U}}_j$ from a MUB, ${\cal M}_{\delta}$, where $\delta = S$,
at each stage of the recursion. More precisely, we seed with ${\cal{U}}_j({\bf y}_j)$
where ${\cal{U}}_j$ is taken from ${\cal M}_{\delta}$, and the variables, ${\bf y}_j$, are
introduced via another matrix. This is made clear in the next subsection.
The complementary properties, (i.e. PAPR $\le S = \delta$)
are guaranteed because every member of ${\cal M}_{\delta}$ is a
unitary matrix. Moreover, the pairwise inner-product between arrays or sequences
generated will be small because $\Delta$ is minimised for ${\cal M}_{\delta}$. At the same time, the
size of the sequence set is maximised because we can choose ${\cal{U}}_j$ to be one of
$\delta + 1$ unitaries (for $\delta$ a prime power). Finally the size of the projected
sequence set can be further
increased by choosing one of $\delta !$ permutations of the rows of the unitaries at each stage of
the recursion.

\vspace{2mm}

For general $S = \delta$ it remains open to develop theoretical
formulae for the size of our constructed array and sequence sets, and for the $\Delta$ value
for the sets. But we do solve these issues for the special case where $S = \delta = 2$, and
this is the
topic of the rest of this paper. Note, however, that our choice of ${\cal M}_2$ allows us
express things in terms of generalised Boolean functions, and this facilitates our theoretical
development. It is likely that such functional
methods do not, in general, extend to $S = \delta > 2$.

\subsection{Seeding with ${\cal M}_2$ to generate complementary pairs}
We now consider the special case where ${\cal{U}}_j$ is selected from
${\cal M}_2 = \{I,H,N\}$ at each stage of the recursion, i.e. we focus on the case where
$\delta = S = 2$, and generate complementary pairs.
Moreover we let $m_j = 1$, $\forall j$, so $\mu_j = j$
and ${\cal{U}}_j({\bf y}_j) = {\cal{U}}_j(z_j)$, $\forall j$.

\vspace{3mm}

From (\ref{GenComp1}), let $F_j'({\bf z}_j) = {\cal{U}}'_j(z_j)F_{j-1}'({\bf z}_{j-1})$ where,
using the equivalence of (\ref{equiv}), we set
${\cal{U}}'_j(z_j) = P_{\theta_{j+1}}{\cal{U}}_j(z_j)P_{\gamma_j}$.
Let $F_j({\bf z}_j) = P^{-1}_{\theta_{j+1}}F_j'({\bf z}_j)$. Then
$$  F_j({\bf z}_j) = {\cal{U}}_j(z_j)P_{\gamma_j}P_{\theta_j}F_{j-1}({\bf z}_{j-1}). $$
Without loss of generality we simplify $P_{\gamma_j}P_{\theta_j}$ to $P_j$, and obtain
$$  F_j({\bf z}_j) = {\cal{U}}_j(z_j)P_jF_{j-1}({\bf z}_{j-1}). $$
We now separate ${\cal{U}}_j(z_j)$ into MUB and variable parts:
$$ {\cal{U}}_j(z_j) = {\cal{U}}_jP_{{\cal U}_j}V_j(z_j), $$
where $P_{{\cal U}_j}$ is a permutation unitary (the diagonal unitary, $O_{{\cal U}_j}$, is
set to $I$ for the reason given in subsection \ref{MUB}),
$V_j(z_j) =
\left ( \begin{tiny} \begin{array}{cc} 1 & 0 \\ 0 & z_j \end{array} \end{tiny} \right )$ and
${\cal{U}}_j \in {\cal M}_2$. So
$$ F_j({\bf z}_j) = {\cal{U}}_jP_{{\cal U}_j}V_j(z_j)P_jF_{j-1}({\bf z}_{j-1}). $$
As $P_{{\cal U}_j} \in \{I,X\}$, $HX = ZH$, and $NX = iZXN$, we
swap ${\cal{U}}_j$ and $P_{{\cal U}_j}$, replace $P_{j+1}P_{{\cal U}_j}$ by
$P_{j}$, and obtain
\beg F_j({\bf z}_j) = P_j{\cal{U}}_jV_j(z_j)F_{j-1}({\bf z}_{j-1}),
\label{MUBrecursion} \eeg
where $P_j \in \{I,X\}$.
(We ignore global constants such as `$i = \sqrt{-1}$' as they have no effect on
our final construction).

\vspace{3mm}

Let
${\cal{U}} = ({\cal{U}}_0,{\cal{U}}_1,\ldots,{\cal{U}}_{n-1}) \in {\cal M}_2^n$.
Then we recurse (\ref{MUBrecursion}) $n$ times so as to construct
\beg
\begin{array}{l}
F_{n-1}({\bf{z}}) = \left ( \begin{array}{l} F_{n-1,0}({\bf{z}}) \\ F_{n-1,1}({\bf{z}}) \end{array} \right ), \\
\hspace{40mm} \m{ where } F_{n-1,k}({\bf{z}}) = c\sum_{{\bf{x}} \in \F_2^n} f_{n-1,k}({\bf{x}}){\bf{z}}^{\bf{x}},
\hspace{7mm} k \in \{0,1\},
\end{array}
\label{MUBResult} \eeg
${\bf{z}}^{\bf{x}} = \prod_{j=0}^{n-1} z_j^{x_j}$, and
$c$ is some real constant such that $F_{n-1,k}({\bf{z}})$ is normalised
as an array (sum of element square-magnitudes is 1). It remains to characterise $f_{n-1,k}$.

\vspace{2mm}

\begin{figure*}[h]
\centering
\includegraphics[height=3in]{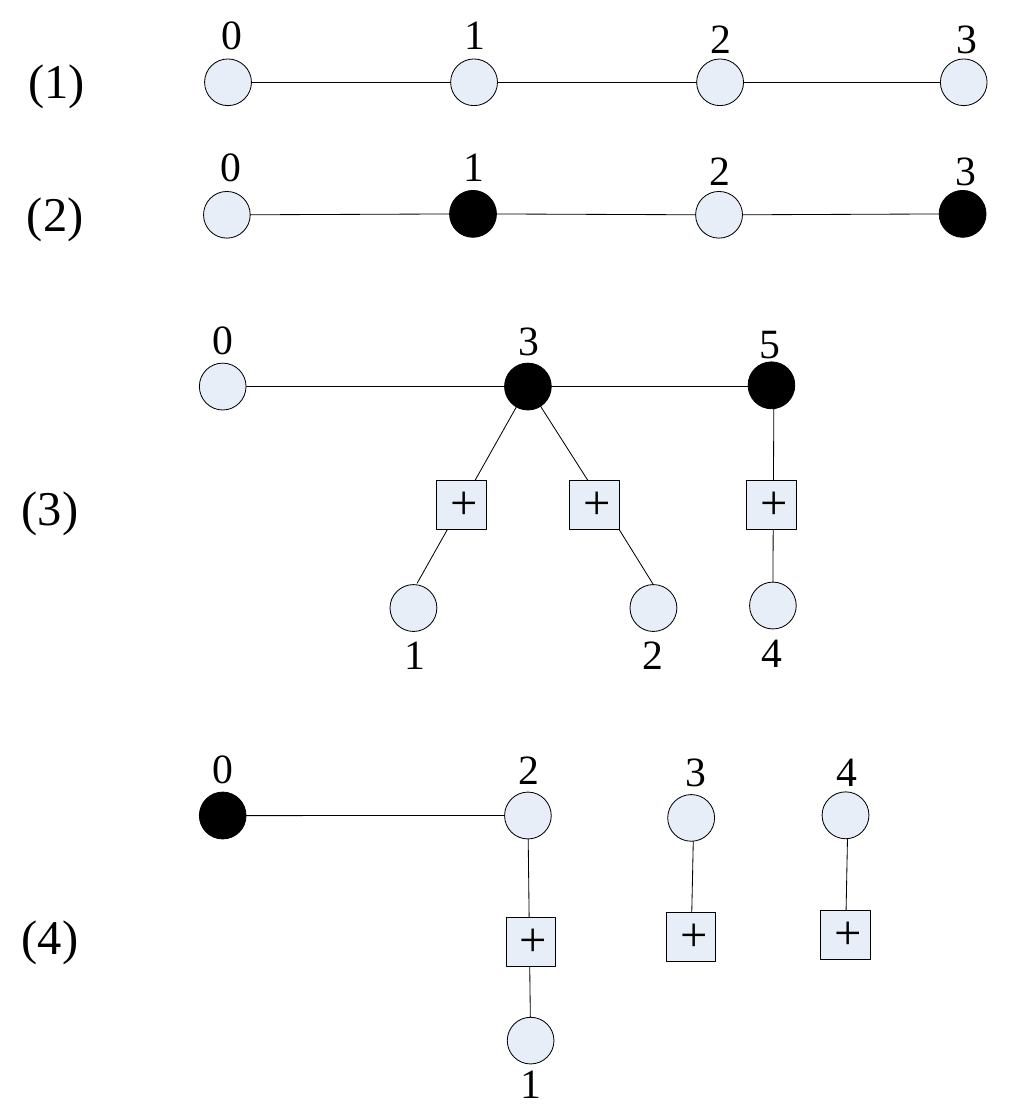}
\caption{Graph Representations of Example Functions}
\label{GraphEx}
\end{figure*}

To begin with, let ${\cal{U}}_j = H$ and $P_j = I$, $\forall j$. Then we construct
$$ f_{n-1,k}({\bf{x}}) : \F_2^n \rightarrow \{1,-1\}
 = i^{2(kx_{n-1} + \sum_{j=0}^{n-2}x_jx_{j+1})}. $$
These are binary complementary sequences, as constructed in \cite{davis}.
This function is illustrated by (1) in Fig \ref{GraphEx}, and is given by
$$ {\cal{U}} = (H,H,H,H) \Rightarrow f_{3,0}({\bf{x}}) = i^{2(x_0x_1 + x_1x_2 + x_2x_3)}. $$

\vspace{2mm}

More generally, let ${\cal{U}}_j \in \{H,N\}$ and let
$l = (j, {\cal{U}}_j = N)$. Then we construct
$$ f_{n-1,k}({\bf{x}}) : \F_2^n \rightarrow \{1,i,-1,-i\}
 = i^{2(kx_{n-1} + \sum_{j=0}^{n-2}x_jx_{j+1}) + \sum_{j=0}^{|l|-1} x_{l(j)}}. $$
These are quaternary complementary sequences, as constructed in \cite{davis}.
An example of this function for $l = (1,3)$ is illustrated by (2) in Fig \ref{GraphEx}, and is
given by
$$ {\cal{U}} = (H,N,H,N) \Rightarrow f_{3,0}({\bf{x}}) = i^{2(x_0x_1 + x_1x_2 + x_2x_3) + x_1 + x_3}. $$

\vspace{2mm}

More generally, let ${\cal{U}}_j \in {\cal M}_2 = \{I,H,N\}$, where
${\cal{U}}_{n-1} \ne I$, and let $p = (j, {\cal{U}}_j \in \{H,N\})$,
$s = (j, {\cal{U}}_j = I)$ and
let $q(v) = j$ if ${\cal{U}}_j \ne I$ and
${\cal{U}}_i = I$, $\forall i$, $v < i < j$, $j < n$, $j \ne v$, and let $q(v) = n$ otherwise. Then we construct
$$ f_{n-1,k}({\bf{x}}) : \F_2^n \rightarrow {\cal A}
 = (\prod_{j=0}^{|s|-1} (x_{s(j)} + x_{q(s(j))} + 1))
i^{2(kx_{p(|p|-1)} + \sum_{j=0}^{|p|-2}x_{p(j)}x_{p(j+1)}) + \sum_{j=0}^{|l|-1} x_{l(j)}}, $$
where $p(-1) = n$, $x_n = 0$, and ${\cal A} = \{0,1,i,-1,-i\}$.
An example of this function for $p = (0,3,5)$, $l = (3,5)$, $s = (1,2,4)$, and
$q = (3,3,3,5,5,6)$,
is illustrated by (3) in Fig \ref{GraphEx}, and is given by
$$ {\cal{U}} = (H,I,I,N,I,N) \Rightarrow
 f_{5,0}({\bf{x}}) = (x_1 + x_3 +1)(x_2 + x_3 +1)(x_4 + x_5 +1)i^{2(x_0x_3 + x_3x_5) + x_3 + x_5}. $$

\vspace{2mm}

More generally, let ${\cal{U}}_j \in {\cal M}_2 = \{I,H,N\}$.
Moreover, if,
for some $t$, ${\cal{U}}_{n-1} = {\cal{U}}_{n-2}=\cdots= {\cal{U}}_{n-t} = I$,
$0 \leq t \leq n$, and
${\cal{U}}_{n-t-1} \neq I$, then define $b$ such that
$b(j) = 1$ for $j \ge n - t$, and $b(j) = 0$ otherwise. Then we construct
(\ref{fMUBprior}). An example of this function for $p = (0,2)$, $l = (0)$, $s = (1,3,4)$,
$q = (2,2,5,5,5)$,
and $b = (0,0,0,1,1)$, is illustrated by (4) in Fig \ref{GraphEx}, and is given by
$$ {\cal{U}}=(N,I,H,I,I) \Rightarrow
    f_{4,k}{\bf(x)}=(x_1+x_2+1)(x_3 + k + 1)(x_4 + k + 1)i^{2(kx_2 + x_0x_2) + x_0}. $$

\vspace{3mm}

\vspace{3mm}

We summarise the previous discussion with the following definition.
\begin{df} Let $p = (j, {\cal{U}}_j \in \{H,N\})$, $l = (j, {\cal{U}}_j = N)$, and
$s = (j, {\cal{U}}_j = I)$ be vectors of integers ordered by magnitude.
Let $q(v) = j$ if ${\cal{U}}_j \ne I$ and
${\cal{U}}_i = I$, $\forall i$, $v < i < j$, $j < n$, $j \ne v$, and let $q(v) = n$ otherwise.
If, for some $t$, ${\cal{U}}_{n-1} = {\cal{U}}_{n-2}=\cdots= {\cal{U}}_{n-t}= I$, $0 \leq t \leq n$, and
${\cal{U}}_{n-t-1} \neq I$, then define $b$ such that $b(j) = 1$ for $j \ge n - t$, and $b(j) = 0$ otherwise.
\label{vecdef}
\end{df}


%

Then we can construct (\ref{MUBResult}), where
\beg f_{n-1,k}({\bf{x}}) : \F_2^n \rightarrow {\cal A}
 = (\prod_{j=0}^{|s|-1} (x_{s(j)} + x_{q(s(j))} + kb(s(j)) + 1))
i^{2(kx_{p(|p|-1)} + \sum_{j=0}^{|p|-2}x_{p(j)}x_{p(j+1)}) + \sum_{j=0}^{|l|-1} x_{l(j)}}, \label{fMUBprior} \eeg
where $p_{-1} = n$, $x_{n} = 0$, and ${\cal A} = \{0,1,i,-1,-i\}$.

\vspace{2mm}




The graphical language of Fig \ref{GraphEx} generalises the path graph
of \cite{davis,paterson} whilst, at the same time,
making a precise mathematical
connection with factor graph notation \cite{Frank}, and to quantum graph states \cite{Hein}
and their generalisations \cite{Riera}.


\vspace{2mm}

As mentioned previously, we interpret the coefficients of $F_{n-1,k}({\bf{z}})$ as an array in
$({\mathbb C}^2)^{\otimes n}$, i.e. an $n$-dimensional $2 \times 2 \times \ldots \times 2$ complex
array. However, from (\ref{MUBResult}), $F_{n-1,k}({\bf{z}})$ is wholly dependent on
$f_{n-1,k}({\bf{x}})$ and, in the following, much of the exposition will be developed in terms of
$f$ rather than $F$, so we also refer to $f$ as an array, where the array elements are the $2^n$
evaluations of $f$ at ${\bf{x}} \in \F_2^n$. The subsequent
projections of $F$, from array to sequence, then carry over to $f$ in an obvious way.

\vspace{3mm}

In (\ref{fMUBprior}), we have, for ease of exposition, set
$P_j = I$, $\forall j$. More generally,
let $r = (r(0),r(1),\ldots,r({n-1})) \in \F_2^n$ be such that $P_j = X^{r(j)}$, $\forall j$.
Moreover, let \\
$w = (w(0),w(1),\ldots,w({n-1})) \in \F_2^n$, where
$$ w(i) = \sum_i^{q(i)-1} r(i). $$
Our later enumerations
multiply by $2^n$ to take into account that $r \in \F_2^n$, and this multiplicity carries over to $w$ as the mapping
from $r \rightarrow w$ is one-to-one.
We then generalise (\ref{fMUBprior}) to:
\beg
\begin{array}{l}
f_{n-1,k}({\bf{x}}) : \F_2^n \rightarrow {\cal A} = \\
 \hspace{10mm} (\prod_{j=0}^{|s|-1} (x_{s(j)} + x_{q(s(j))} + kb(s(j)) + w({s(j)}) + 1)) \times \\
\hspace{34mm} i^{2(kx_{p(|p|-1)} + \sum_{j=0}^{|p|-1} w({p(j)})x_{p(j)} +
\sum_{j=0}^{|p|-2}x_{p(j)}x_{p(j+1)}) + \sum_{j=0}^{|l|-1} x_{l(j)}},
\end{array} \label{FullEq} \eeg

From (\ref{FullEq}), the set of complementary arrays is
${\cal B}_n = \{f_{n-1,0}({\bf{x}}) | {\cal U} \in {\cal M}_2^n, r \in F_2^n\}$. In the next section we evaluate
$|{\cal B}_n|$ where, evidently, $|{\cal B}_n| \le 6^n$.

\vspace{3mm}

The arrays constructed in Theorem 1 of \cite{bparker09} are
a subset of those described by (\ref{FullEq}), corresponding to the case where
${\cal U} \in \{I,H\}^n$, with $t = 0$. \cite[Conjecture 1]{bparker09} can, consequently,
be improved to
\begin{conj}
For any $n$, each type-I complementary array over the alphabet $\{0,1,-1\}$
is of the form
\beg
f_{n-1,0}({\bf{x}}) \\
 \hspace{2mm} = (\prod_{j=0}^{|s|-1} (x_{s(j)} + x_{q(s(j))} + w({s(j)}) + 1))
(-1)^{\sum_{j=0}^{|p|-1} w({p(j)})x_{p(j)} +
\sum_{j=0}^{|p|-2}x_{p(j)}x_{p(j+1)}}. \label{fMUBconj}
\eeg
\end{conj}
(`Type-I' complementarity is just the form of complementarity discussed in this paper for
arrays over $({\mathbb C}^2)^{\otimes n}$).

\section{Enumerations}\label{enum}
In this section we evaluate $|{\cal B}_n|$. We also evaluate $|{\cal B}_{\downarrow,n}|$, which is the number
of complementary sequences of length $2^n$ that can be obtained from arrays in ${\cal B}_n$ by the
projections, $z_i = z^{2^{\pi(i)}}$, $\pi \in {\cal S}_n$,
from the $n$-dimensional arrays down to $1$ dimensional sequences of length $2^n$. Clearly there are
an infinite number of other projections one could choose, e.g.
$z_i = z^{3^{\pi(i)}}$, $\pi \in {\cal S}_n$, and for which complementarity would be preserved, but we
do not consider such variations in this paper.

\subsection{Number of arrays in ${\cal B}_n$}
We associate ${\cal U}$ with a length $n$ binary sequence $a=(a_0,a_1,\cdots, a_{n-1})$, where
$0$ represents $I$ and $1$ represents $H $ or $N$.
Let
$$ {\cal B'}_n = \{ {\cal B}_n |{\cal U}_{n-1} \neq I \}. $$

\begin{thm}\label{arrayno}
\bearray
|{\cal B'}_n| =\left\{
\begin{aligned}
&2^n\cdot \sum_{k=0}^{n-1}(\binom{n-1}{k}2^{n-k-1}+\binom{\frac{n}{2}-1}{\lfloor\frac{k}{2}\rfloor}\cdot
2^{\lceil\frac{n-k}{2}\rceil -1} )=2^n\cdot 3^{n-1}+2^{n+1}\cdot 3^{\frac{n}{2}-1}, \ \mbox{for}\  n \mbox{ even}, \\
&2^n\cdot \sum_{k=0}^{n-1}(\binom{n-1}{k}2^{n-k-1}+\binom{\frac{n-1}{2}}{\frac{k}{2}}\cdot
2^{\lceil\frac{n-k}{2}\rceil -1} )=2^n\cdot 3^{n-1}+2^{n}\cdot 3^{\frac{n-1}{2}}, \ \mbox{for}\  n \mbox{ odd}, \\
\end{aligned} \right. \label{arraynumber}
 \eearray
where  $\binom{n}{t}=0, $ if $t$ is not an integer.
\end{thm}

\pf
There are
$\binom{n-1}{k}$ binary sequences of length $n-1$ with $k$ zeros. Let  $\binom{n-1}{k}=S_k+A_k$,
where $S_k$ (or $A_k$)
is the number of symmetric (or asymmetric) length $n-1$ binary sequences with $k$ zeros. We have that
 $$
S_k =\left\{
\begin{aligned}
&\binom{\frac{n-1}{2}}{\frac{k}{2}}, \ \mbox{for}\  n-1 \mbox{ even}, \\
&\binom{\lfloor\frac{n-1}{2}\rfloor}{\lfloor\frac{k}{2}\rfloor}, \ \mbox{for}\  n-1 \mbox{ odd}. \\
\end{aligned} \right.
$$
Let $2^{n-k}=S'_k+A'_k,$ where $S'_k$ (or $A'_k$) is the number of symmetric (or  asymmetric)
binary sequences of length $n-k$.
Note that $S'_k=2^{\lceil\frac{n-k}{2}\rceil}$ and $a_{n-1}=1$.
Let $a$ have $k$ zeros. Then, when $(a_0,a_1,\cdots,a_{n-2})$ is symmetric,
there are
$S'_k+\frac{A'_k}{2}=\frac{2S'_k+A'_k}{2}=\frac{2^{n-k}+S'_k }{2}$ choices for the $n-k\ 1's$
to be $H \mbox{ or }N$. When $(a_0,a_1,\cdots,a_{n-2})$ is asymmetric
there are
$2^{n-k}$ choices for the $n-k\ 1's$ to be $H \mbox{ or }N$.
Then
\bearray
|{\cal B'}_n| &=&
2^n\cdot \sum_{k=0}^{n-1}(S_k\cdot \frac{2^{n-k}+S'_k}{2}+\frac{N_k}{2}\cdot 2^{n-k}) \nonumber\\
&=&2^n\cdot \sum_{k=0}^{n-1}((S_k+N_k)\cdot \frac{2^{n-k}}{2}+\frac{S_k}{2}\cdot S'_k) \nonumber\\
&=&2^n\cdot \sum_{k=0}^{n-1}(\binom{n-1}{k}\cdot {2^{n-k-1}}+\frac{S_k}{2}\cdot S'_k) \nonumber\\
&=&\left\{ \begin{aligned}
&2^n\cdot \sum_{k=0}^{n-1}(\binom{n-1}{k}2^{n-k-1}+\binom{\frac{n}{2}-1}{\lfloor\frac{k}{2}\rfloor}\cdot
2^{\lceil\frac{n-k}{2}\rceil -1} ), \ \mbox{for}\  n \mbox{ even}, \\
&2^n\cdot \sum_{k=0}^{n-1}(\binom{n-1}{k}2^{n-k-1}+\binom{\frac{n-1}{2}}{\frac{k}{2}}\cdot
2^{\lceil\frac{n-k}{2}\rceil -1} ), \ \mbox{for}\  n \mbox{ odd}, \\
\end{aligned} \right.
 \eearray \hfill{$\Box$}

\begin{cor}
The number of arrays in ${\cal B}_n$ is
\bearray
|{\cal B}_n| &=&
\sum_{m=0}^{n}
|{\cal B'}_m|
\cdot 2^{n-m}\nonumber\\
&=&\left\{
\begin{aligned}
&2^{n-1}\cdot (3^n+3\cdot 3^{\frac{n}{2}}-2), \ \mbox{for}\  n \mbox{ even}, \\
&2^{n-1}\cdot (3^n+5\cdot 3^{\frac{n-1}{2}}-2), \ \mbox{for}\  n \mbox{ odd}, \\
\end{aligned} \right.
 \eearray where
$|{\cal B'}_0| = 1$.
\end{cor}

In \cite{davis} the authors construct the set of standard quaternary complementary sequences, ${\cal DJ}_n$,
being $\Z_4$-linear offsets
of the $\F_2$ `path graph' \cite{paterson}. Using our terminology this translates to the construction of
(\ref{FullEq}) under the restriction ${\cal U} \in \{H,N\}^n$, i.e. where $|s| = 0$.
Although \cite{davis} only viewed their
objects as complementary sequences they are, more generally, complementary
arrays over $({\mathbb C}^2)^{\otimes n}$
\cite{parker,parker3,Mats,JP,fiedler1,parker2011}. As $n$ gets large, $|{\cal B}_n|$ approaches $6^n$, whereas the
size of the construction of \cite{davis} approaches $4^n$. The larger size of our set is achieved by
enlarging the alphabet from $\{1,i,-1,-i\}$ in \cite{davis} to $\{0,1,i,-1,-i\}$ in this paper, more
accurately, by selecting our unitaries from ${\cal M}_2 = \{I,H,N\}$ instead of from
the sub-optimal MUB $\{H,N\}$. (A sub-optimal MUB is a set of mutually unbiased bases where the
number of bases is less than the maximum possible. When $\delta$ is a prime power, then the maximum
possible number of bases is $\delta + 1$ and, for ${\cal M}_2$, $\delta + 1 = 3$.) But,
crucially, as discussed later, the increase in set size for the set of sequence projections
is achieved without any increase in $\Delta$, i.e.
$\Delta({\cal B}_{\downarrow,n}) = \Delta({\cal DJ}_n)$.


\vspace{2mm}

\subsection{An algorithm for generating all sequences in ${\cal B}_{\downarrow,n}$, and the
corresponding enumeration}
Many practical applications of our construction would exploit the length $2^n$
sequences obtained from the arrays in ${\cal B}_n$ by projection. Such sequences comprise the
set ${\cal B}_{\downarrow,n}$. By projection we mean the following. We have that
$F_{n-1,0}({\bf z}) = c \sum_{{\bf x} \in \F_2^n} f_{n-1,0}({\bf x}){\bf z}^{\bf x}$, for
$f_{n-1,0}$ as defined in (\ref{FullEq}). The
coefficients of this polynomial form an $n$-dimensional array, and can be projected
down to a 1-dimensional array by the assignments $z_j = z^{2^j}$. Such a projection
produces a polynomial in $z$ of degree $2^n - 1$ whose coefficients form a length $2^n$
sequence with elements from ${\cal A} = \{0,1,i,-1,-i\}$. More generally, for each
$F_{n-1,0}({\bf z})$, we generate the
$n!$ projections obtained by assigning $z_j = z^{2^{\pi(j)}}$, $\forall \pi \in {\cal S}_n$,
where ${\cal S}_n$ is the group of permutations of $n$ objects. These
projections can be obtained by generating
${\cal B}_{\downarrow,n} = \{f_{n-1,0}({\bf x}_{\pi}), \forall {\cal U} \in {\cal M}_2^n, r \in \F_2^n, \forall \pi \in {\cal S}_n\}$,
where
${\bf x}_{\pi} = (x_{\pi(0)},x_{\pi(1)},\ldots,x_{\pi(n-1)})$.
Not all these projections are unique when taken over all polynomials in ${\cal B}_n$.
We sketch out, in a stepwise fashion, a recursive algorithm that generates
sequences in ${\cal B}_{\downarrow,n}$ uniquely, firstly when ${\cal U} \in \{I,H\}^n$, and then
for ${\cal U} \in {\cal M}_2^n$. In each case we implemented the algorithm, obtained computational
results, and then proved the results. It should be noted that the theoretical developments were
greatly helped by us first plugging our computational results into the
On-Line Encyclopedia of Integer Sequences (OEIS) \cite{Sloane}.

We refer to ${\cal U} \in \{I,H\}^n$ as an `$IH$ string' and
${\cal U} \in {\cal M}_2^n$ as an `$IHN$ string'. Moreover we abbreviate vectors, e.g.
$(I,H,I,H,H)$ is shortened to $IHIHH$, and $(N,H,H,I,N,I)$ to $NHHINI$, and abbreviate
$f_{n-1,0}({\bf{x}})$ to $f({\bf{x}})$ or $f$.

\subsubsection{An algorithm for ${\cal U} \in \{I,H\}^n$}
We set $P_j = I$, $\forall j$, and choose ${\cal U} \in \{I,H\}^n$.
It is possible that two $IH$ strings generate the same sequence, so
we must consider three uniqueness criteria, as follows.

\vspace{2mm}

\textbf{Criterion A:}
Consider, as an example, that we generate the sequence associated with
${\cal U} = HIIHIHII$, i.e. we generate
$f = (x_1 + x_3 + 1)( x_2 + x_3 + 1)(x_4 + x_5 + 1)(x_6 + 1)(x_7 + 1)(-1)^{x_0x_3 + x_3x_5}$.
The projection down to a sequence
depends on the ordering, $\pi$, of the 8 variables in ${\bf{x}}$, i.e.
$\pi({\bf{x}}) = (x_{\pi(0)},x_{\pi(1)},x_{\pi(2)},x_{\pi(3)},x_{\pi(4)},x_{\pi(5)},
x_{\pi(6)},x_{\pi(7)})$, $\pi \in {\cal S}_8$.
The number of sequences generated in this way is $8!$.

The form of $f$ implies that $x_1 = x_2 = x_3$ and $x_4 = x_5$. Moreover
$x_6$ and $x_7$ can be swapped without changing the function.
So permuting $1$, $2$, and $3$, and/or swapping $4$ and $5$, and/or swapping $6$ and $7$,
has no effect on the function, and reduces the function enumeration to $\frac{8!}{3!2!2!}$.
In order to avoid
these repetitions we only allow $\pi \in {\cal S}_8$ under the conditions
$\pi(1) < \pi(2) < \pi(3)$, $\pi(4) < \pi(5)$, and $\pi(6) < \pi(7)$.

\vspace{2mm}

\textbf{Criterion B:}
Consider the sequence generated by $IHIIHHII$.
This generates
$f' = (x_0 + x_1 + 1)(x_2 + x_4 + 1)(x_3 + x_4 + 1)(x_6 + 1)(x_7 + 1)(-1)^{x_1x_4 + x_4x_5}$.
For the example of $f$ discussed for criterion A,
$f(\pi({\bf{x}})) = f'({\bf{x}})$ for some $\pi \in {\cal S}_8$. This is because
$IHIIH$ is the reversal of $HIIHI$. So, to ensure unique generation,
only one of $f$ or $f'$ should be generated. Ignoring the rightmost $HII$, one
can interpret the $IH$ strings, $HIIHIHII$ and $IHIIHHII$, as binary strings
$10010$ and $01001$, respectively.
With the least significant bit on the left, we equate these strings with integers 9 and 18, respectively, and
throw away, arbitrarily, the $IH$ string associated with the largest number,
namely $IHIIHHII$.

\vspace{2mm}

\textbf{Criterion C:}
The symmetry of criterion B does not occur if the $IH$ substring is symmetric under reversal.
For instance, consider the string $IHHIHI$. Then, ignoring the rightmost $HI$, we see that
$IHHI$ is symmetric. So there is no $f'$ to throw away. However, this
symmetric condition leads, instead, to an alternative restriction on the allowed permutation $\pi$.
For instance, for this example, we allow only one of the permutations
$(354120)$ and $(124350)$, (which are both valid under previous conditions on $\pi$)
as they both lead to
$f = (x_1 + x_2 + 1)(x_3 + x_5 + 1)(x_0 + 1)(-1)^{x_2x_4 + x_4x_5}$. Ignoring all integers to
the right of the position of the rightmost $H$, i.e. in this case ignoring `$0$', we
choose, arbitrarily,
to throw away the permutation with the lowest integer on the right-hand side - in this case
we throw away
$(354120)$ as `$1$' is on the right-hand side of $(35412)$.
One needs to refine this decision process. Consider the string $IHIHIHI$, and permutations
$(4602351)$ and $(3502461)$, which are both valid under previous conditions on $\pi$.
We see that $IHIHI$ is symmetric.
Then, ignoring `$1$', the lowest integer, `$0$', is in the centre of $(460235)$, as $x_0 = x_2$,
so we choose to decide between the two permutations on the next lowest off-centre integer. In this
case, we decided based on integer `$3$' and throw away $(4602351)$ as `$3$' is right of centre
in this permutation.

\vspace{5mm}

We have implemented a recursive algorithm based on criteria $A$, $B$, and $C$, and
obtain the enumerations shown in Table \ref{IHNoLin}. Let us call this number $E_{IH}(n)$.
We show that $E_{IH}(n)$ is sequence A032262 of \cite{Sloane}.

\begin{table}\label{IHNoLin}
\begin{small}
\begin{tabular}{|c|cccccccccc|} \hline
$n$                    & 1 &  2   &  3   & 4   & 5   & 6    & 7     & 8      & 9       & 10          \\ \hline
$E_{IH}(n)$            & 2 & 5    & 17   & 83  & 557 & 4715 & 47357 & 545963 & 7087517 & 102248075   \\ \hline
${\tilde E}_{IH}(n)$   & 2 & 6    & 26   & 150 & 1082& 9366 & 94586 & 1091670& 14174522& 204495126   \\ \hline
$\log_2(E_{IH}(n))$    & 1 & 2.32 & 4.09 & 6.38& 9.12& 12.20& 15.53 & 19.06  & 22.76   & 26.61       \\ \hline
$\log_2(\frac{n!}{2})$ & 0 & 0    & 1.58 & 3.58& 5.91&  8.49& 11.30 & 14.30  & 17.47   & 20.79       \\ \hline
\end{tabular}
\caption{$E_{IH}(n) = \#$Unique $IH$ Sequences, $P_j = I$, $\forall j$ - A032262\cite{Sloane}}
\end{small}
\end{table}

\vspace{3mm}

We make use of the following combinatoric numbers:
$$ \begin{array}{l}
\text{\em Stirling's number of the second kind: } \\
S_2(n,k) = \left \{ \begin{array}{cc} n \\ k \end{array} \right \} =
\frac{1}{k!} \sum_{j=0}^k (-1)^{k-j}\binom{k}{j}j^n. \end{array} $$

\beg  \begin{array}{l} \text{\em generalized ordered Bell numbers: } \\
B(r,n) = r\sum_{k=1}^n \binom{n}{k}B(r,n-k)
 = \sum_{k=0}^n r^kk! \left \{ \begin{array}{c} n \\ k \end{array} \right \}, \mf B(r,0) = 1.
 \end{array} \label{BRecursion} \eeg
$B(r,n)$ is A094416 of \cite{Sloane}.
The case when $r = 1$ generates the ordered Bell numbers.

\begin{thm}
The enumeration of $IH$ strings of length $n$, taking into account criteria A, B, and C, is given by
$$ E_{IH}(n) = 2^{n-1} + \sum_{k=0}^n k! \left \{ \begin{array}{c} n \\ k \end{array} \right \}. $$
\label{EIH}
\end{thm}

\begin{pf} (of theorem \ref{EIH})
We first enumerate ${\tilde E}_{IH}(n)$, which only takes criterion A into account.
Ignoring criteria B and C causes a double count of unique $IH$ strings except for
$III\ldots I$, $HII \ldots I$, $IHI \ldots I$, $IIH \ldots I$, \ldots, $III \ldots H$,
i.e. the exceptions are all $IH$ strings of $H$ weight less than 2, for which neither
the symmetry of criterion B or of C is possible. These exceptions
contribute an additive
correction factor of $2^n$ to ${\tilde E}_{IH}(n)$. We, thereby, obtain a relationship
between $E_{IH}(n)$ and ${\tilde E}_{IH}(n)$:
\beg {\tilde E}_{IH}(n) = 2E_{IH}(n) - 2^n.
\label{AOnly} \eeg

We show that ${\tilde E}_{IH}(n)$ is A000629 of \cite{Sloane}, and
include ${\tilde E}_{IH}(n)$ in table \ref{IHNoLin}.

\vspace{3mm}

\begin{thm}
The enumeration of $IH$ strings, based only on criterion A, is given by
$$ {\tilde E}_{IH}(n) = 2B(1,n), \mf n > 0, \mf {\tilde E}_{IH}(0) = 1. $$
\label{IHEnum}
\end{thm}
\par {\addtolength{\leftskip}{5mm}
\begin{pf} (of theorem \ref{IHEnum})
Consider all $IH$ strings of length $n-1$ of the form $\ldots H$, i.e. with a
rightmost $H$. Taking into account only criterion A, then
let us say that there are $B(n-1)$ unique strings of this type. Let $s_{n-1}$ be any such $IH$ string of length $n-1$.
Now consider all $IH$ strings of length $n$ of the form $s_{n-1}I = \ldots HI$, i.e. with a rightmost $HI$. Considering
all variable index permutations, $\pi \in {\cal S}_n$, the single rightmost $I$ can be associated with one of $n$ indices. So there are
$\binom{n}{1}B(n-1)$ unique strings of the form $s_{n-1}I = \ldots HI$. More generally, consider
all $IH$ strings of length $n-k$ of the form $\ldots H$. There are $B(n-k)$ unique strings of this type.
Let $s_{n-k}$ be any such $IH$ string of length $n-k$, and
consider all $IH$ strings of length $n$ of the form $s_{n-k}II\ldots I = \ldots HII\ldots I$, i.e. with $k$
rightmost $I$'s. Considering
all variable index permutations, $\pi \in {\cal S}_n$, the $k$ rightmost $I$s can be associated with $\binom{n}{k}$ indices.
So there are
$\binom{n}{k}B(n-k)$ unique strings of the form $s_{n-k}II\ldots I$, i.e. with $k$ rightmost $I$s.
With initial conditions $B(0) = 1$, we have that the number of unique $IH$ strings of length $n$ and
with at least one rightmost $I$, taking into account only criterion A, is given by
\beg B'(n) = \sum_{k=1}^n \binom{n}{k}B(n-k), \mf B(0) = 1. \label{RightI} \eeg

We are left with enumerating the number, $B(n)$, of unique $IH$ strings with a rightmost $H$, taking into account only
criterion A. We find that

\beg B(n) = B'(n). \label{BB} \eeg
\par {\addtolength{\leftskip}{10mm}
\begin{pf} (of (\ref{BB}))
Let $s_n = vH$ and $t_n = vI$ be two $IH$ strings of length $n$ with a rightmost $H$ and $I$, respectively, and where
$v$ is an $IH$ string of length $n-1$. Let $v$ have $r$ rightmost $I$s. Then, using criterion A, permutation of the rightmost $r+1$
indices of $s_n$ is a symmetry. Similarly, permutation of the rightmost $r+1$ indices of $t_n$ is also a symmetry. It follows that
the enumeration for $IH$ strings with a rightmost $H$ is identical to that for $IH$ strings with a rightmost $I$.
\end{pf}
\par }

\vspace{2mm}

\noindent Theorem \ref{IHEnum} follows from (\ref{RightI}) and (\ref{BB}).
\end{pf} 
\par }

\vspace{2mm}

\noindent Combining (\ref{BRecursion}), for $r = 1$, with theorem \ref{IHEnum} and (\ref{AOnly}) yields
theorem \ref{EIH}.
\end{pf} 

\vspace{3mm}

An asymptotic formula for $E_{IH}(n)$ can be derived from known results on the
asymptote of ordered Bell numbers \cite{Barthelemy,Wilf} and
A000670 of \cite{Sloane}:
\beg E_{IH}(n)_{n \rightarrow \infty} = \frac{n!}{2} \log_2(e)^{n+1}. \label{EIHAsymp} \eeg

\vspace{4mm}

In table \ref{IHNoLin} we also compare $\log_2(E_{IH})$ with $\log_2$ of the number of binary
standard Golay sequences, where $P_j = I$, $\forall j$ (i.e. ignoring linear offsets).

\subsubsection{The $IHN$ strings}
Having sketched out an algorithm to generate all sequences uniquely from $IH$ strings, and
derived associated enumeration formulae, we now
extend to $IHN$ strings. Consider $HIIHIHII$. If we now
include the possibility
of $N$ then, once we have generated $HIIHIHII$,
along with a specified permutation, we must also generate $NIIHIHII$, $HIINIHII$, \\
$NIINIHII$, $HIIHINII$, $NIIHINII$, $HIININII$, and $NIININII$, i.e. 8 $IHN$ strings in total, all
with the same permutation. Criteria $A$, $B$, and $C$ have already been tackled in generating the
initial $IH$ string, so need not be re-considered.

\vspace{5mm}

We have extended our recursive algorithm to $IHN$ strings, and
obtain the enumerations shown in Table \ref{IHNNoLin}. Let us call this number $E_{IHN}(n)$,
where $|{\cal B}_{\downarrow,n}| = 2^nE_{IHN}(n)$. The $2^n$ factor occurs because we must, more generally,
consider $P_j \in \{I,X\}$, $\forall j$.

\begin{table}\label{IHNNoLin}
\begin{small}
\begin{tabular}{|c|ccccccccc|} \hline
$n$                  & 1  &  2   &  3   & 4   & 5   & 6    & 7     & 8      & 9         \\ \hline
$E_{IHN}(n)$         & 3  &11    & 63   &563  &6783 &99971 &1724943&34031603&755385183  \\ \hline
${\tilde E}_{IHN}(n)$& 3  & 15   & 111  & 1095&13503&199815&3449631&68062695&1510769343 \\ \hline
$\log_2(E_{IHN}(n))$ &1.58& 3.46 & 5.98 & 9.14&12.73&16.61 & 20.72 & 25.02  & 29.49     \\ \hline
$\log_2(n!2^{n-1})$  & 0  & 2    & 4.58 & 7.58&10.91& 14.49& 18.30 & 22.30  & 26.47     \\ \hline
\end{tabular}
\caption{$E_{IHN}(n) = \#$Unique $IHN$ Sequences, $P_j = I$, $\forall j$}
\end{small}
\end{table}

\begin{thm}
The enumeration of $IHN$ strings of length $n$, taking into account criteria A, B, and C, is given by
$$ E_{IHN} = 3\sum_{k=0}^n 2^{k-2}k!\left \{ \begin{array}{c} n \\ k \end{array} \right \}
 + 2^n - \frac{1}{2}. $$
\label{EIHN}
\end{thm}

\begin{pf}  (of theorem \ref{EIHN})
We first enumerate ${\tilde E}_{IHN}(n)$, which only takes criterion A into account.
Ignoring criteria B and C causes a double count of unique $IHN$ strings except
those with $H$ or $N$ of weight less than 2, which contribute an additive
correction factor of $2^{n+1} - 1$ to ${\tilde E}_{IHN}(n)$. We, thereby, obtain a relationship
between $E_{IHN}(n)$ and ${\tilde E}_{IHN}(n)$:
\beg {\tilde E}_{IHN}(n) = 2E_{IHN}(n) - 2^{n+1} + 1.
 \label{AOnlyIHN} \eeg
We show that ${\tilde E}_{IHN}(n)$ is
A201339 of \cite{Sloane}, and include ${\tilde E}_{IHN}(n)$ in table \ref{IHNNoLin}.

\vspace{3mm}

\begin{thm}
The enumeration of $IHN$ strings, based only on criterion A, is given by
$$ {\tilde E}_{IHN}(n) = \frac{3}{2}B(2,n), \mf n > 0, \mf {\tilde E}_{IHN}(0) = 1. $$
\label{IHNEnum}
\end{thm}
\par {\addtolength{\leftskip}{5mm}
\begin{pf}   (of theorem \ref{IHNEnum})
Let $R \in \{H,N\}$.
Consider all $IHN$ strings of length $n-k$ of the form $\ldots R$. Taking into account criterion A, let us
say that there are $C(n-k)$ unique strings of this type.
Let $s_{n-k}$ be any such $IHN$ string of length $n-k$, and
consider all $IHN$ strings of length $n$ of the form $s_{n-k}II\ldots I = \ldots RII\ldots I$,
i.e. with $k$ rightmost $I$'s. Considering
all variable index permutations, $\pi \in {\cal S}_n$, the $k$ rightmost $I$s can be associated with $\binom{n}{k}$ indices.
So there are
$\binom{n}{k}C(n-k)$ unique strings of the form $s_{n-k}II\ldots I$, i.e. with $k$ rightmost $I$s.
With initial conditions $C(0) = 1$, we have that the number of unique $IHN$ strings of length $n$ and
with at least one rightmost $I$, taking into account only criterion A, is given by
\beg C'(n) = \sum_{k=1}^n \binom{n}{k}C(n-k), \mf C(0) = 1. \label{IHNRightI} \eeg

We are left with enumerating the number, $C(n)$, of unique $IHN$ strings with a rightmost $H$ or $N$,
taking into account only criterion A. We find that
\beg C(n) = 2C'(n). \label{CC} \eeg
\par {\addtolength{\leftskip}{10mm}
\begin{pf} (of (\ref{CC}))
Let $s_n = vR$ and $t_n = vI$ be two $IHN$ strings of length $n$ with a rightmost $H$ or $N$, and $I$, respectively, and where
$v$ is an $IHN$ string of length $n-1$. Let $v$ have $r$ rightmost $I$s. Then, using criterion A, permutation of the rightmost $r+1$
indices of $s_n$ is a symmetry. Similarly, permutation of the rightmost $r+1$ indices of $t_n$ is also a symmetry,
irrespective of whether the rightmost element is $H$ or $N$. It follows that
the enumeration for $IHN$ strings with a rightmost $H$ or $N$ is exactly twice that for $IH$ strings with a rightmost $I$.
\end{pf}
\par }

\vspace{2mm}

\noindent Theorem \ref{IHNEnum} follows from (\ref{IHNRightI}), (\ref{CC}), and (\ref{BRecursion}).
\end{pf}
\par }

\vspace{2mm}

\noindent Combining (\ref{BRecursion}), for $r = 2$, with theorem \ref{IHNEnum} and (\ref{AOnlyIHN})
yields theorem \ref{EIHN}.
\end{pf} 

\vspace{3mm}

An asymptotic formula for $E_{IHN}(n)$ can be derived from known results on the asymptote
for $B(2,n)$ \cite{Cloitre}:
\beg E_{IHN}(n)_{n \rightarrow \infty} = \frac{n!}{4\ln (\frac{3}{2})^{n+1}}. \label{EIHNAsymp} \eeg

\vspace{4mm}

In table \ref{IHNNoLin} we also compare $\log_2(E_{IHN})$ with $\log_2$ of the number of $\Z_4$
standard Golay sequences, where $P_j = I$, $\forall j$ (i.e. ignoring linear offsets).

\subsubsection{The $IHN$ strings plus binary linear offsets}
Once we have generated the $IHN$ strings uniquely, the more general choice of $P_j \in \{I,X\}$,
$\forall j$,
replaces $E_{IHN}(n)$ with $|{\cal B}_{\downarrow,n}| = 2^nE_{IHN}(n)$, so all values in table \ref{IHNNoLin} are multiplied by $2^n$
($n$ is added to all log values).

\section{Pairwise inner-product}\label{distance}
In this section we consider $\Delta^2({\cal B}_{\downarrow,n})$, where
$$ \Delta^2({\cal B}_{\downarrow,n})=\mbox{ max}\{\Delta^2(f,f')|f \neq f',
f,f' \in {\cal B}_{\downarrow,n} \}. $$

It is worth re-iterating that each $f$ represents one of $n!$ sequences, obtained from
each $F({\bf z}) = c\sum_{\bf x} f({\bf x}){\bf z}^{\bf x}$ by projection to $F(z)$, where
the sequence, $F$, is formed from the coefficients of $F(z)$. Moreover the normalising constant, $c$,
is chosen so that $\|F\|^2 = \sum_{0 \le i < n} F_{i}F_{i}^* = 1$ (see (\ref{FullEq}) for more details).

\begin{thm}\label{InnerP}
$\Delta^2({\cal B}_{\downarrow,n})=\frac{1}{2}$.
\end{thm}

\pf Let $f, f'\in {\cal B}_{\downarrow,n}$. Assume that $f$ and  $f'$  have  $2^{n-e}$ and $2^{n-e'}$
elements which are not zero, respectively.  Then
$h=f \cdot f' $ has $2^{n-u}$ elements which are not
zero, where $\mbox{ max }\{e,e'\}\leq u\leq e+e'$. We obtain the bound
\beg \Delta^2(f,f') \le \frac{(2^{n-u})^2}{2^{n-e}2^{n-e'}} = 2^{e+e'-2u}. \label{OverlapBound} \eeg
There are three cases, where the bound of (\ref{OverlapBound}) suffices for cases 2 and 3.

\noindent \textbf{Case 1: $e=e',$ and $f,f'$ have the same positions that
are non-zero.}
 Let $ \widehat{f}$ and $\widehat{f'}$
be  the compressed length $2^{n-e}$ sequences obtained by deleting the zeros in
$f$ and $f'$, respectively.
 Let $ \widehat{f}=i^{\tilde{f}},  \widehat{f'}=i^{\tilde{f'}},$
where
 $\tilde{f},\tilde{f'} : \F_2^{n-e} \rightarrow \Z_4$  are generalized Boolean functions.
 If the linear terms of $ \tilde{{f}}$ and $\tilde{{f'}}$
 are the same, then the Hamming distance between  $ \tilde{{f}}$ and $\tilde{{f'}}$
 is at least $2^{n-e-2}$,
 and $2^{n-e-2}\leq N_2\leq 3\cdot 2^{n-e-2}$, $2^{n-e-2}\leq N_0\leq 3\cdot 2^{n-e-2}$,
 $N_1=N_3=0$,  where
$$ N_j = |\{{\bf x}|\tilde{{f}}({\bf x})- \tilde{{f'}}({\bf x})=j,{\bf x}\in F_2^{n-e}\}|, \hspace{5mm} j \in \Z_4.$$
Then
$$|\langle f,f' \rangle|^2=|\sum_{{\bf x}\in F_2^{n-e}}i^{\tilde{{f}}({\bf x})- \tilde{{f'}}({\bf x})}|^2
=|N_2-N_0|^2+|N_3-N_1|^2\leq |2^{n-e-1}|^2,$$ and
  $$ \Delta^{2}( f, f')=\frac{|\langle f,f' \rangle|^2}{2^{n-e}\cdot2^{n-e}}
\leq\frac{|2^{n-e-1}|^2}{2^{n-e}\cdot2^{n-e}}=\frac{1}{4}.$$

 If the linear terms of $ \tilde{{f}}$ and $\tilde{{f'}}$
 are different then,
 since $ (\tilde{{f}}- \tilde{{f'}}) \mbox{ mod }2$
is a balanced Boolean function over $\Z_2$, then
$N_0+N_2=N_1+N_3=2^{n-e-1}$.
 Then $$|\langle f,f' \rangle|^2=|\sum_{{\bf x}\in F_2^{n-e}}i^{\tilde{{f}}({\bf x})- \tilde{{f'}}({\bf x})}|^2
=|N_2-N_0|^2+|N_3-N_1|^2\leq 2\cdot |2^{n-e-1}|^2,$$
$$ \Delta^{2}(f,f')=\frac{|\langle f,f' \rangle|^2}{2^{n-e}\cdot2^{n-e}}
\leq\frac{2\cdot |2^{n-e-1}|^2}{2^{n-e}\cdot2^{n-e}}=\frac{1}{2}.$$

\noindent \textbf{Case 2: $e= e'$, but the non-zero positions of  $f$ and  $f'$
are different.} Then $e+1\leq u$ and, from (\ref{OverlapBound}),
$$ \Delta^{2}( f,f') \le \frac{1}{4}. $$

\noindent \textbf{Case 3: $e\neq e'$.} Wlog assume $e<e'$. Then $e'\leq u$, and, from (\ref{OverlapBound}),
$$ \Delta^{2}( f,f') \le \frac{1}{2}. $$

It remains to exhibit a pair of sequences
 $a,a' \in {\cal B}_{\downarrow,n}$ satisfying $\Delta^{2}(a,a')=\frac{1}{2}$.
 For example, $a=(-1)^{\sum_{i=0}^{n-2}x_ix_{i+1}}, a'= a \cdot i^{x_{n-1}}$.
 \hfill{$\Box$}

For the set, ${\cal DJ}_n$, of standard quaternary Golay sequences,
$\Delta^{2}({\cal DJ}_n)=\frac{1}{2}$, because
 ${\cal DJ}_n \subset {\cal B}_{\downarrow,n}$ and $a,a' \in  {\cal DJ}_n$.


\section{A codebook from a subset of ${\cal B}_{\downarrow,n}$}\label{codebooks}
In this section we give a construction for a codebook, ${\mathcal C}$, over ${\cal A}$
that is a subset of ${\cal B}_{\downarrow,n}$.
The maximum magnitude of inner products between distinct codewords is approximately
$\sqrt{\frac{3}{2}}$ times the Welch bound for large $n$.

An $({\cal N},{\cal K})$ codebook,
$\mathcal{C}=\{\textbf{c}_0,\textbf{c}_1,\cdots, \textbf{c}_{{\cal N}-1}\}$,
is a set of ${\cal N}$ distinct codewords in a ${\cal K}$-dimensional vector space where
${\cal K} \leq {\cal N}.$ Each code vector
$\textbf{c}_h=(c_{h,0},c_{h,1},\cdots, c_{h,{\cal K}-1}),0\leq h < {\cal N},$
has unit-norm, i.e., $\|\textbf{c}_h\|=\sqrt{\sum_{i=0}^{{\cal K}-1}|c_{h,i}|^2}=1$.
Welch \cite{welch} gave a well-known  lower bound on $\Delta(\mathcal{C}):$
$$ \Delta(\mathcal{C})=\mbox{max}_{0\leq h\neq m < {\cal N}}|\textbf{c}_h\textbf{c}_m^{\dag}|\geq
\Delta_{welch}(\mathcal{C})=\sqrt{\frac{{\cal N}-{\cal K}}{{\cal K}({\cal N}-1)}}. $$
If $\Delta(\mathcal{C}) = \Delta_{welch}(\mathcal{C})$, then
$\mathcal{C}$ is called a maximum-Welch-bound-equality (MWBE) codebook.


%

Abbreviate $f_{n-1,0}(\bf x)$ by $f$, let $f = f_{\cal U}$
for some fixed ${\cal U} \in {\cal M}_2^n$, and construct the codeset
$$ {\mathcal C}_{\cal U} = \{f_{\cal U} \hspace{2mm} | \hspace{2mm} r \in \F_2^n\}. $$
Then ${\mathcal C}_{\cal U}$ comprises $2^n$ pairwise orthogonal sequences.

\vspace{2mm}

\noindent Let ${\cal R} = \{{\cal R}_0,{\cal R}_1,{\cal R}_2\}$,
${\cal R}_j = ({\cal R}_{j,0},{\cal R}_{j,1},\ldots,{\cal R}_{j,n-1}) \in {\cal M}_2^n$, $0 \le j \le 2$,
such that
\beg {\cal R}_{1,i} = I \m{ iff } {\cal R}_{0,i} \ne I, \mf 0 \le i < n. \label{cond1} \eeg
Let $p_j$, $j \in \{0,1\}$, be the vector, $p$, as defined in Definition \ref{vecdef},
for ${\cal U} = {\cal R}_j$,
with border conditions $p_0(-1) = p_1(-1) = -1$. We also require auxiliary vectors
$u = {\cal R}_0 \cdot {\cal R}_1 \in \{H,N\}^n$ and $w \in \Z^n$. We construct $w$ and
${\cal R}_2$ element-by-element, i.e. $w_i$ then ${\cal R}_{2,i}$ then
$w_{i+1}$ then ${\cal R}_{2,i+1}$, etc, starting with $w_{p_j(i_j)} = w_0$, where
\beg \begin{array}{l}
w_{p_j(i_j)} = |\{h \mz | \mz {\cal R}_{2,p_j(i_j-1)+h} = N, \mf
 0 < h < p_j(i_j) - p_j(i_j-1)\}|, \mz 0 \le i_j < |p_j|, j \in \{0,1\}, \\
\begin{array}{lll}
{\cal R}_{2,i} & = H & \m{ iff } (w_i \m{ odd }, u(i) = H) \m{ or } (w_i \m{ even }, u(i) = N), \\
              & = N & \m{ otherwise.}
\end{array}
\end{array} \label{makeodd} \eeg

For instance, if ${\cal R}_0 = (I,I,N,I,I,H,I,I)$ and  ${\cal R}_1 = (H,N,I,H,H,I,N,N)$ then
$p_0 = (2,5)$, $p_1 = (0,1,3,4,6,7)$, and $u = (H,N,N,H,H,H,N,N)$.
Then
$$ \begin{array}{ll}
w_{p_1(0)} = w_0 = |\{h \mz | \mz {\cal R}_{2,p_1(0-1)+h} = N,  \mf 0 < h < 0-(-1) = 1\}| = 0 &
             {\cal R}_{2,0} = N \\
w_{p_1(1)} = w_1 = |\{h \mz | \mz {\cal R}_{2,p_1(1-1)+h} = N,  \mf 0 < h < 1-0 = 1\}| = 0 &
             {\cal R}_{2,1} = H \\
w_{p_0(0)} = w_2 = |\{h \mz | \mz {\cal R}_{2,p_0(0-1)+h} = N,  \mf 0 < h < 2-(-1) = 3\}| = 1 &
             {\cal R}_{2,2} = N \\
w_{p_1(2)} = w_3 = |\{h \mz | \mz {\cal R}_{2,p_1(2-1)+h} = N,  \mf 0 < h < 3-1 = 2\}| = 1 &
             {\cal R}_{2,3} = H \\
w_{p_1(3)} = w_4 = |\{h \mz | \mz {\cal R}_{2,p_1(3-1)+h} = N,  \mf 0 < h < 4-3 = 1 \}| = 0 &
             {\cal R}_{2,4} = N \\
w_{p_0(1)} = w_5 = |\{h \mz | \mz {\cal R}_{2,p_0(1-1)+h} = N,  \mf 0 < h < 5-2 = 3 \}| = 1 &
             {\cal R}_{2,5} = H \\
w_{p_1(4)} = w_6 = |\{h \mz | \mz {\cal R}_{2,p_1(4-1)+h} = N,  \mf 0 < h < 6-4 = 2 \}| = 0 &
             {\cal R}_{2,6} = H \\
w_{p_1(5)} = w_7 = |\{h \mz | \mz {\cal R}_{2,p_1(5-1)+h} = N,  \mf 0 < h < 7-6 = 1 \}| = 0 &
             {\cal R}_{2,7} = H.
\end{array} $$
So $w = (0,0,1,1,0,1,0,0)$ and ${\cal R}_2 = (N,H,N,H,N,H,H,H)$.
As another example, if ${\cal R}_0 = (H,I,I,I,N)$ and  ${\cal R}_1 = (I,H,H,H,I)$ then
$p_0 = (0,4)$, $p_1 = (1,2,3)$, $u = (H,H,H,H,N)$. Then $w = (0,1,0,0,3)$,
and ${\cal R}_2 = (N,N,N,N,N)$.


The codebook, ${\mathcal C}$, is then constructed as
\beg {\mathcal C}  = {\mathcal C}_{{\cal R}_0} \cup {\mathcal C}_{{\cal R}_1}
 \cup {\mathcal C}_{{\cal R}_2} = \{f_{{\cal U}} \hspace{2mm} | \hspace{2mm}
 {\cal U} \in {\cal R}, r \in \F_2^n\}.
\label{codebookgen} \eeg
${\mathcal C}$ is actually a codebook of arrays $\in ({\mathcal C}^2)^{\otimes n}$, being
$\subset {\cal B}_n$, but we further view ${\mathcal C}$ as a codebook of sequences $\subset {\cal B}_{\downarrow,n}$
by subsequent projections, as discussed previously.

\vspace{2mm}

\begin{thm}
${\mathcal C}$, as constructed in (\ref{codebookgen}), is a $(3\times 2^n,2^n)$ codebook, where
$\Delta({\mathcal C})=\sqrt{2^{-n}}$, $\Delta_{welch}({\mathcal C})
 = \sqrt{\frac{1}{3.2^n - 1}}$,
$\Delta({\mathcal C}) \rightarrow \sqrt{\frac{3}{2}}\Delta_{welch}({\mathcal C})$, as
$n \rightarrow \infty$.
\label{cbook} \end{thm}
\pf
Let $p_j$, $s_j$, $l_j$, and $r_j$ be the vectors $p,s,l,r$, respectively, for
${\cal U} = {\cal R}_j$, as defined in  Definition \ref{vecdef}.
(\ref{cond1}) implies that $f_{{\cal R}_0}$ and $f_{{\cal R}_1}$ are both nonzero at
only one element, for any $r_0,r_1 \in \F_2^n$.
Moreover $\|f_{{\cal R}_j}\| = \sqrt{2^{n - |p_j|}}$, and $p_0 + p_1 = n$.
It follows that $\Delta(f_{{\cal R}_0},f_{{\cal R}_1}) = \sqrt{2^{-n}}$.

\noindent For $\Delta(f_{{\cal R}_0},f_{{\cal R}_2})$ and $\Delta(f_{{\cal R}_1},f_{{\cal R}_2})$ we require
the following identity:

\vspace{2mm}

\noindent For any $a \in \F_2^n$,
\beg \begin{array}{lll}
|\sum_{x \in \F_2^n} i^{2a \cdot x + b \cdot x}|^2 & = 2^n, &
     \mf b = (1,1,\ldots,1)^T.
\end{array} \label{AllZ4} \eeg
\pf (of (\ref{AllZ4})) is straightforward and is omitted. \hfill{$\Box$}

From (\ref{FullEq}),
$f_{n-1,0}$ can be written as
$$ f_{n-1,0}({\bf x}) = \chi({\bf x})i^{{\cal P}({\bf x})}, $$
where $\chi({\bf x}) \in \F_2^n$ is a product of linear constraints, and
${\cal P}({\bf x}) = 2{\cal Q}({\bf x}) + {\cal L}({\bf x})$ is the sum of a binary quadratic term,
${\cal Q}$, and a $\Z_4$-linear term, ${\cal L}$. Let
$f_{{\cal R}_j} = \chi_j({\bf x})i^{{\cal P}_j({\bf x})}$, where
${\cal P}_j = 2{\cal Q}_j + {\cal L}_j$.

To prove for $\Delta(f_{{\cal R}_0},f_{{\cal R}_2})$ we first set $r_0 = r_2 = 0$. Then
\beg \begin{array}{ll}
f_{{\cal R}_0}f_{{\cal R}_2}
 & = \chi_0({\bf x})i^{{\cal P}_0({\bf x})}\chi_2({\bf x})i^{{\cal P}_2({\bf x})}
 = \chi_0({\bf x})i^{{\cal P}_0({\bf x})}i^{{\cal P}_2({\bf x})} \mf \m{ (as $\chi_2 = 1$)} \\
 & = \chi_0({\bf x})i^{{\cal P}_0({\bf x})}\chi_0({\bf x})i^{{\cal P}_2({\bf x})} \\
 & = \chi_0({\bf x})i^{2{\cal Q}_0({\bf x}) + {\cal L}_0({\bf x})}
     i^{2({\cal Q}_0({\bf x}) + {\cal L}'({\bf x})) + {\cal L}_2({\bf x})} \mf \m{ for some binary linear $L'$} \\
 & \hspace{60mm} \m{(as $\chi_0$ restricts ${\cal Q}_2$ to ${\cal Q}_0 + {\cal L}'$)} \\
 & = \chi_0({\bf x})i^{2{\cal L}'({\bf x}) + {\cal L}_0({\bf x}) + {\cal L}_2({\bf x})}.
\end{array} \label{QOut} \eeg
The key point in (\ref{QOut}) is that the quadratic terms, ${\cal Q}_0$, cancel.
$\chi_0$ equates subsets of the variables in ${\bf x}$ as follows:
\beg y_i = x_{p_0(i-1) + 1} = x_{p_0(i-1) + 2} = \ldots = x_{p_0(i)},  \mf 0 \le i < |p_0|. \label{restrict} \eeg
We re-express (\ref{QOut}) as
$$ f_{{\cal R}_0}f_{{\cal R}_2}
 = i^{2{\cal L}'({\bf y}) + {\cal L}_0({\bf y}) + {\cal L}_2({\bf y})}, \mf {\bf y} \in \F_2^{|p_0|}. $$
Then
$$ \langle f_{{\cal R}_0},f_{{\cal R}_2} \rangle
 = \sum_{{\bf y} \in \F_2^{|p_0|}} i^{2{\cal L}'({\bf y}) + {\cal L}_0({\bf y}) + {\cal L}_2({\bf y})}. $$
It follows, using (\ref{AllZ4}), that
$|\langle f_{{\cal R}_0},f_{{\cal R}_2} \rangle |^2 = 2^{|p_0|}$ iff
\beg {\cal L}_0({\bf y}) + {\cal L}_2({\bf y}) = 2{\cal L}''({\bf y}) + {\bf 1} \cdot {\bf y},
\label{lincond} \eeg
for some binary linear term, ${\cal L}''$. Our construction satisfies (\ref{lincond}) because the
conditions of (\ref{makeodd}) on ${\cal R}_2$
ensure the contribution of an odd number of $N$ terms for each $y_i$, i.e. the addition
of an odd number of $\Z_4$ linear terms, $y_i$, $\forall i$,
$0 \le i < |p_0|$, giving $y_i$ or $3y_i = 2y_i + y_i$, where $2y_i$ contributes to ${\cal L}''$.
It follows that $\Delta^2(f_{{\cal R}_0},f_{{\cal R}_2})
 = \frac{|\langle f_{{\cal R}_0},f_{{\cal R}_2} \rangle |^2}{2^{|p_0|}2^n} = 2^{-n}$.

The generalisation to any $r_0,r_1 \in F_2^n$ simply adds more binary linear terms to ${\cal L}'$
in (\ref{QOut}) and/or changes $x_{p_0(i-1) + h}$ to $x_{p_0(i-1) + h} + 1$ for one or more $h$. Neither
modification affects the result.

The proof for $\Delta(f_{{\cal R}_1},f_{{\cal R}_2})$ is identical to that for
$\Delta(f_{{\cal R}_0},f_{{\cal R}_2})$. \hfill{$\Box$}

\vspace{2mm}

\begin{lem}
There are $2^{n-1}(2^n - 1)$ distinct codebooks, ${\cal C} \in {\cal B}_n$, of arrays that can be generated by (\ref{codebookgen}).
\end{lem}
\pf (sketch) There are $2^n$ ways to share out the $I$'s between ${\cal R}_0$ and ${\cal R}_1$. For each one
of these, there are $2^n$ choices for $\{H,N\}^n$. But ${\cal R}_0$ and ${\cal R}_1$ could be swapped so, to avoid this
symmetry, set the first element of ${\cal R}_0$ equal to $I$, which halves our count so we get $2^{2n-1}$. However this
includes the possibility that ${\cal R}_0 = (I,I,I,\ldots,I)$, in which case ${\cal R}_1,{\cal R}_2 \in \{H,N\}^n$ and there
is a double count as ${\cal R}_1$ and ${\cal R}_2$ could be swapped, so we must subtract $2^{n-1}$ from $2^{2n-1}$.
\hfill{$\Box$}

\vspace{2mm}

We leave open the problem of enumerating the number of distinct codebooks, ${\cal C} \in {\cal B}_{\downarrow,n}$,
of sequences, as generated by (\ref{codebookgen}).

\vspace{2mm}

We can further construct a codebook that is a subset of ${\mathcal C}$ and that approaches
$\sqrt{2}\Delta_{welch}$ for large $n$, as follows.

\vspace{2mm}

Let ${\cal R} = \{{\cal R}_0,{\cal R}_1\}$,
${\cal R}_j \in \{I,H\}^n$, $0 \le j \le 1$,
and let
\beg {\mathcal C}^{IH}  = \{f_{{\cal U}} \hspace{2mm} | \hspace{2mm}
 {\cal U} \in {\cal R}, r \in \F_2^n\},
\label{codebookgenIH} \eeg
under the condition
$$ \begin{array}{ll}
{\cal R}_{1,i} = I \m{ iff } {\cal R}_{0,i} \ne I, \mf 0 \le i < n.
\end{array} $$

\begin{thm}
${\mathcal C}^{IH}$ is a $(2^{n+1},2^n)$ codebook, where
$\Delta({\mathcal C}^{IH}) \rightarrow \sqrt{2}\Delta_{welch}({\mathcal C}^{IH})$, as
$n \rightarrow \infty$.
\label{cbookIH} \end{thm}
\pf Omitted - a simple subcase of the previous construction. \hfill{$\Box$}



\vspace{3mm}

By mapping $I \rightarrow 0$, $N \rightarrow 1$, $H \rightarrow 2$, one can interpret
${\cal R}$ as a code over $\F_3$, i.e. a ternary code. For instance, one choice of
${\cal R}$ is $\{II\ldots I, NN\ldots N, HH\ldots H\}$ which can be interpreted as the ternary
repetition code $\{00\ldots 0, 11\ldots 1, 22\ldots 2\}$ so, in a sense, we are constructing a codebook
from a ternary code. It would be interesting, in future work, to construct codebooks
from other, larger, ternary codes, both linear and nonlinear.

\vspace{2mm}

In \cite{Nam} a codebook that asymptotes to $\sqrt{6}$ times the Welch bound is constructed.
It is interesting because of the way it is constructed from complementary sequences, and is motivated, in
particular, by application to compressed sensing. However, unlike ${\cal B}_{\downarrow,n}$,
the codebook
members are not designed to satisfy an upper bound on PAPR, so the codebook of \cite{Nam} is
not comparable in this sense to ${\cal B}_{\downarrow,n}$.

\section{Conclusion}\label{conclusion}
We have described a construction for complementary sets of arrays and sequences using a recursive
matrix notation, and then proposed to seed the construction with an optimal MUB. Specifically, we focused
on the ${\cal M}_2$ case to construct complementary pairs. Thereby, we
constructed a set, ${\cal B}_n$, of
complementary arrays over $({\mathbb C}^2)^{\otimes n}$, and have
projected ${\cal B}_n$
down to a set, ${\cal B}_{\downarrow,n}$, of length $2^n$ complementary sequences
that is a superset of the quaternary set of standard complementary sequences, ${\cal DJ}_n$, constructed in \cite{davis}.
Whilst $|{\cal B}_{\downarrow,n}|$ is significantly larger than $|{\cal DJ}_n|$, the
PAPR upper bound remains at 2, and the magnitude of the pairwise inner-product between set members
remains at $\frac{1}{\sqrt{2}}$, i.e. $\Delta({\cal B}_{\downarrow,n}) = \Delta({\cal DJ}_n)$.

Unlike most constructions for QAM complementary sequences in the literature, the sequences
in ${\cal B}_{\downarrow,n}$
all possess the same power, i.e. $\|F,F\|^2 = 1$, $\forall F \in {\cal B}_{\downarrow,n}$.
So the upper bound on PAPR for
the sequence carries over to the upper bound on PAPR for the set. The QAM constructions in
the literature typically
propose sets of sequences with varying powers - this could be a disadvantage in some
applications. This `equal power'
property for ${\cal B}_{\downarrow,n}$ would carry over to complementary constructions
using more general ${\cal M}_{\delta}$.
However, this `equal power' property also has a downside when ${\cal M}_2$ is used.
For example, consider the size $2^n$
subset of ${\cal B}_{\downarrow,n}$ sequences constructed using ${\cal U} = (I,I,I, \ldots, I)$, $r \in \F_2^n$.
These are $2^n$ spikes or pulses of very large relative magnitude and, in some contexts, it may be
practically undesirable to generate such sequences. So, in some contexts, one might
choose to generate the subset of ${\cal B}_{\downarrow,n}$,
generated from ${\cal U} \in {\cal M}_2^{\otimes n}$, where the number of $I$'s in ${\cal U}$
is not too big.

\vspace{2mm}

We also extracted a codebook from ${\cal B}_{\downarrow,n}$ that achieves $\sqrt{\frac{3}{2}}$ times the
Welch bound for large $n$. It is of interest because every member of the codebook
also satisfies PAPR $\le 2$.

\vspace{2mm}

The primary aim of this paper is to advertise the central role played by the optimal MUB in our
construction. Although the method is applicable to an optimal MUB of any dimension, we focused on
${\cal M}_2 = \{I,H,N\}$,
of dimension $\delta = 2$, so as to recursively construct the complementary sequence set,
${\cal B}_{\downarrow,n}$. The parameters
of ${\cal M}_2$ control the parameters of ${\cal B}_{\downarrow,n}$ in that the sequences of
${\cal B}_{\downarrow,n}$ satisfy a PAPR upper-bound of $\delta = 2$ precisely because $I$, $H$, and
$N$ are $\delta \times \delta$ unitary, the value of
$|{\cal B}_{\downarrow,n}|$ is large because $|{\cal M}_2| = 3 = \delta + 1$, the maximum value
possible, and the value of $\Delta({\cal B}_{\downarrow,n})$ is small because
$\Delta = \frac{1}{\delta} = \frac{1}{2}$ for ${\cal M}_2$ is the minimum possible.

\subsection{Some open problems}
When $\delta$ is a prime power then we know
that $|{\cal M}_{\delta}| = \delta + 1$. If, in (\ref{GenComp1}), we assign the non-variable part
of $\{{\cal U}_j({\bf y}_j)\}$ to ${\cal M}_{\delta}$ then we generalise the results of this paper to
$\delta \ge 2$. One expects the complementary set of arrays, ${\cal B}_{{\cal M}_{\delta},n}$, and sequences,
${\cal B}_{{\cal M}_{\delta},\downarrow,n}$, constructed from the recursion of (\ref{GenComp1}),
to have very good properties.
We know that the PAPR for ${\cal B}_{{\cal M}_{\delta},n}$ and ${\cal B}_{{\cal M}_{\delta},\downarrow,n}$,
is upper-bounded by $\delta$, but
\bite
    \item Can one develop expressions for $|{\cal B}_{{\cal M}_{\delta},n}|$
    and $|{\cal B}_{{\cal M}_{\delta},\downarrow,n}|$
    in terms of just $\delta$ and $n$?
    \item Can one develop expressions for $\Delta({\cal B}_{{\cal M}_{\delta},\downarrow,n})$
    in terms of just $\delta$ and $n$?
\eite

The use of ${\cal M}_2$ in our construction ensures that our arrays and sequences have elements drawn from
the alphabet $\{0,1,i,-1,-i\}$ (up to normalisation), and this facilitates our development of expressions
for $|{\cal B}_{\downarrow,n}|$ and $\Delta({\cal B}_{\downarrow,n})$. There exist similar optimal Fourier-based MUBs, ${\cal M}_{\delta}$, for
$\delta$ a prime power, where the functional approach used in this paper seems to generalise naturally. But not all optimal MUBs are of this form so,
for general MUBs, we propose the following challenge:
\bite
    \item Design an algorithm for the receiver, based on the matrix alphabet, ${\cal M}_{\delta}$, to decode received sequences
    that have been generated at the transmitter using the complementary construction seeded by ${\cal M}_{\delta}$.
\eite

In section \ref{codebooks}, we suggest that it is somewhat
natural to map from the ternary alphabet $\{0,1,2\}$ to
${\cal M}_2 = \{I,N,H\}$. Thus `strong' codes over $\{0,1,2\}$ might be used to
select relatively strong subsets of ${\cal B}_n$ and
${\cal B}_{\downarrow,n}$. It would be interesting to investigate, not just the ternary case, but more general mappings from codes over
$\{0,1,\ldots,\delta\}$ to subsets of the complementary array/sequence sets that have been seeded using ${\cal M}_{\delta}$.

We constructed a codebook that meets $\sqrt{\frac{3}{2}}$ times the Welch bound as $n \rightarrow \infty$.
Although the codebook does not meet the Welch bound with equality, it does have the extra rather
strict constraint that every sequence in the set satisfies PAPR $\le 2$. This extra constraint is
well-motivated for, for instance, spread-spectrum applications, and suggests the following
challenge:
\bite
    \item We wish to find an infinite construction for a codebook, ${\cal C}$,
    where every sequence in ${\cal C}$ satisfies PAPR $\le T$. Then how close can ${\cal C}$
    get to the Welch bound? One expects that the answer depends on $T$. The problem also clearly
    depends on the relative sizes of ${\cal K}$ and ${\cal N}$. For instance, let ${\cal C}$
    be the subset of ${\cal B}_{\downarrow,n}$ where ${\cal U} = (H,H,\ldots,H)$ and $r \in \F_2^n$.
    Then ${\cal N} = |{\cal C}| = 2^n$, ${\cal K} = 2^n$, so the Welch bound is $0$ and is met
    with equality by ${\cal C}$ as the sequences in ${\cal C}$ are pairwise orthogonal. Moreover
    PAPR$({\cal C}) \le 2$. So the interesting cases occur for ${\cal K} > {\cal N}$.
\eite

\vspace{2mm}

Finally, we have focused on recursive complementary constructions seeded by optimal MUBs such as ${\cal M}_2$. More generally,
it would be interesting to seed our construction with other unitary matrix sets that are not necessarily MUBs. Moreover, we could even seed with
sets of non-unitary matrices, in which case we could obtain larger sets at the price of PAPR
rising with $n$, and larger $\Delta$. For instance, it would be interesting
to seed with the single $\delta^2 \times \delta$ matrix whose $\delta^2$ rows
comprise an equiangular tight frame of dimension $\delta$,
where $\Delta$ for the frame is $\frac{1}{\delta+1}$.

\vspace{5mm}

\noindent {\bf Acknowledgement: } We wish to thank the reviewers for their helpful comments on the paper.

\end{document}